\documentclass[preprint,flushrt]{aastex}
\usepackage{natbib}

\slugcomment{Draft of \today}

\begin{document}

\title{A Study of 3CR Radio Galaxies from z = 0.15 to 0.65.  II. 
Evidence for an Evolving Radio Structure}

\author{Michael Harvanek\altaffilmark{1} \& John T. Stocke}
\affil{Center for Astrophysics and Space Astronomy, CB 389, University 
of Colorado, Boulder, Colorado,
80309-0389, \\ electronic mail: harvanek@apo.nmsu.edu, 
stocke@casa.colorado.edu}

\altaffiltext{1}{Current Address: Apache Point Observatory, 2001 Apache
Point Road, P.O. Box 59, Sunspot, NM, 88349-0059}


\begin{abstract}

Radio morphology data have been collected for a sample of radio 
galaxies from the Revised 3rd Cambridge (3CR)
Catalog in the redshift range
$0.15 < z < 0.65$.  Radio structure parameters including largest 
physical size, projected bending angle
($\beta$),  lobe length asymmetry ($Q$) and hot spot placement 
(Fanaroff-Riley ratio) have been measured from
the highest quality  radio maps available.   Combined with similar 
data for quasars in the same redshift
range, these morphology data are used in conjunction with a 
quantification  of the richness of the cluster
environment around these objects (the amplitude of the galaxy-galaxy 
spatial covariance function,
$B_{gg}$) to search for indirect evidence of a dense intracluster 
medium (ICM). This is accomplished by
searching for confinement and distortions of the radio structure that 
are correlated with $B_{gg}$.

Correlations between physical size and hot spot placement with
$B_{gg}$ show evidence for an ICM  only at $z \leq 0.4$,  but there 
are no correlations  at $z \geq 0.4$,
suggesting an epoch of $z \sim 0.4$ for the formation of the ICM in 
these Abell richness class 0-1, FR2-selected clusters.  X-ray
selected clusters at comparable redshifts, which contain FR1 type
sources exclusively, are demonstrably richer than the FR2-selected
clusters found in this study.
The majority of the radio sources with high
$B_{gg}$ values at $z\leq$0.4 can be described as ``fat doubles'' or 
intermediate FR2/FR1s. The lack of
correlation between $B_{gg}$ and
$\beta$  or
$B_{gg}$ and $Q$ suggests that these types of radio source distortion 
are caused by something other than
interaction with a dense ICM.  Therefore, a large $\beta$ cannot be 
used as an unambiguous indicator of a rich
cluster  around powerful radio sources. These results support the 
hypothesis made in Paper 1 that cluster
quasars fade to become FR2s,  then FR1s, on a timescale of 0.9 Gyrs 
(for H$_0=$ 50 km s$^{-1}$ Mpc$^{-1}$).

\end{abstract}

\keywords{galaxies: active --- galaxies: clusters: general --- 
galaxies: evolution --- intergalactic medium
--- quasars: general --- radio continuum: galaxies}

\section{Introduction} Just as the solar wind was first detected 
indirectly using comet tails, the
intracluster medium (ICM) was first inferred to be present due to the 
morphology of extended radio sources
associated with clusters. Fiducial studies of these cluster radio 
sources date from the 1970s with the work of
\cite{Owen}, \cite{OwenRud} and \cite{BOR79}.
While some details concerning the exact 
relationship between the radio  galaxy
and the ICM which surrounds it are still the subject of some debate 
\citep{Roettiger, Eilek}, all
current models for cluster radio source distortion invoke a dense 
ICM, in an amount consistent with its direct
detection via thermal X-ray bremsstral\"ung \citep{Sarazin}.  At 
low-$z$ ($\leq$0.1) there is a good
correlation (but with a large dispersion)  between the density of the 
ICM (as measured by its X-ray emission)
and the cluster galaxy density \citep[e.g.,][]{AK83, YE02}. 
However, since galaxies formed before the ICM,
and at least partially created it through internal galaxy processes 
(i.e., supernovae and stellar winds), a
cluster with a high galaxy density may not yet have formed a dense 
ICM at its observed epoch.

While sensitive X-ray observations have discovered and studied the 
ICM of  rich clusters out to $z\approx$1
\citep[e.g.,][]{GL94,Donahue,Rosati},  direct X-ray detection of a 
cluster ICM around radio galaxies and
quasars is made much more difficult because the AGN itself is a 
strong X-ray emitter. Because quasars and
powerful radio galaxies are found in clusters only  at higher 
redshifts (Hill \& Lilly 1991, HL hereafter; Ellingson, Yee \& Green
1991, EYG hereafter; Harvanek et al. 2001, Paper 1 hereafter), the 
presence or absence of a dense ICM
surrounding luminous AGN in clusters has not been generally 
established through direct ICM detection (although
a few recent detections have been made; see below). Thus, the use of 
the radio source morphology to trace
indirectly  the presence (or absence) of a cluster ICM is still an 
important indirect method.

In general, radio sources associated with AGN can be divided into two 
basic types based upon morphology and
radio power level: Fanaroff-Riley Type 1 and Type 2 
\citep[FR1 and FR2;][]{FR}. The distorted morphology of the FR1 type radio galaxies 
indirectly reveals the presence of a
dense ICM in current epoch clusters. On the other hand, quasars, 
which are exclusively FR2 type sources, and
powerful FR2 radio galaxies are found in clusters only at
$z\geq$0.2 (HL and Paper 1).  The FR classes were originally 
defined using the distance between the
brightest flux points on opposite sides of the radio core divided by 
the total extent of the source measured
from the faintest radio contour. If this ratio of distances (called 
herein the $FR$ ratio) is $<$  0.5, the
source is classified as an FR1 type. If the $FR$ ratio $>$  0.5, the 
source is an FR2. Qualitatively, FR2s
have a relatively weak radio core with an extended ``lobe'' of 
emission on each side.  The lobes tend to be
fairly symmetric in both size and luminosity and are usually 
collinear (i.e., both lobes and the core lie on
approximately the same line; although see
Stocke, Burns \& Christiansen 1985
for examples of ``bent'' FR2s).   The brightest regions 
of these sources tend to occur at or near
the leading edges of the radio lobes and often these sources contain 
a weak, one-sided jet.  In contrast, FR1s
are asymmetric and distorted and may bear little resemblance to a 
double-lobed structure.  These sources tend
to have bright cores and/or bright, two-sided jets and the extended 
structure of the source dims with distance
from the core. FR2s tend to have a larger physical size than FR1s. 
Thus, an FR1 type structure is small,
distorted, asymmetric, edge-dimmed and (core+jet)-dominated.  FR2 type 
structures tend to be larger, collinear,
symmetric, edge-brightened and lobe-dominated. Additionally, sources 
with 178 MHz radio power
$P_{178} \lesssim 2 \times 10^{25}$ W Hz$^{-1}$ sr$^{-1}$ are FR1s 
while those with higher power are FR2s
\citep{FR}. More recent work on the FR1/FR2 dividing line has found a 
correlation between the radio power level of the dividing line and
the optical host galaxy luminosity, such that more luminous optical 
galaxies can host more luminous FR1 type
sources \citep{OwenLaing}.

While the exact relationship between FR1s and FR2s remains uncertain, 
the work of EYG and Paper 1 has found
evidence that cluster quasars evolve into radio galaxies by fading 
to become first FR2s, then FR1s. This
``fading AGN'' or ``evolutionary'' hypothesis  accounts for the 
presence of  quasars in moderately rich
clusters at $z\sim$0.5 and FR2s in similar richness clusters at
$z\sim$0.25; whereas only FR1s are found in such clusters in the 
current epoch. The ``e-fading'' timescale of
the optical continuum emission from the AGN core suggested by EYG and Paper 1 
is $\sim$0.9 Gyrs (H$_0$=50 km s$^{-1}$ Mpc$^{-1}$).  Since the
timescales associated with extended radio source outbursts are
thought to be
$\sim$10$^8$ yrs
\citep{revmodphys}, the radio source power and morphology will 
``track'' the fading of the AGN. By this
hypothesis, the ICM surrounding the AGN host plays a significant role 
in this process (EYG and Paper 1) by
affecting the triggering and/or fueling of the AGN \citep{Roos,
StocPerr}.
This scenario also explains why
quasars are found in clusters only at $z\geq$0.4 (EYG) and only in 
poorer environments at lower-$z$. However, this
hypothesis remains controversial, especially since another hypothesis
\citep{Barthel} relates quasars and FR2 radio galaxies entirely by orientation.

Indeed, there is much support for an ``obscuration based unification
scheme'', as advocated by Barthel and others \citep[e.g.,][]{Antonucci} in
which quasars and radio galaxies are related not by evolution but by
viewing angle; i.e., quasars are seen closer to their outflow axis than
radio galaxies, causing radio galaxies to be preferentially lower in
optical luminosity due to obscuring material perpendicular to the outflow
axis. Evidence cited in favor of this hypothesis includes: (1). quasars 
have systematically smaller (factor of two at a $\sim90\%$ confidence level) 
extended radio structures on 
the plane of the sky than radio galaxies at the same redshift and radio 
power levels \citep{Barthel}; (2). most luminous 
radio galaxies have only narrow emission lines in their optical spectra 
while all quasars
have broad permitted lines \citep{Antonucci}; (3). some narrow-line radio
galaxies have broad permitted lines in polarized light 
\citep{Antonucci}, including Cygnus A \citep{ogle} (4). the luminosity of 
extended [OII] emission is 
comparable for radio galaxies and quasars with comparable radio power levels 
\citep{Hes}; and (5). the far-infrared dust emission has comparable
luminosities in quasars and radio galaxies at $z>0.8$ \citep{meisenheimer}. 
But none
of these results are without contradictory (or at least confusing) results 
from other investigators: (1). 
many studies on the extended radio size of 
quasars and radio galaxies have been conducted \citep[see][and references
therein]{Urry} with little agreement. 
It may be that the redshift range used
by \citep{Barthel} ($0.5<z<1.0$) is the only redshift range for which this
test gives results in agreement with the unified scheme \citep{Singal};
(2 \& 3). some narrow line radio galaxies at lower redshift have only very weak
emission lines \citep[e.g., the ``low-excitation'' group of][]{laing94}
and would be very unlikely to be ``hidden quasars''. 
The absence of luminous emission line gas
in many FR2s could be an evolutionary effect since FR1s also have no broad line gas, 
very little narrow line gas \citep{owen96} and no evidence for obscuration \citep{falcke}; 
(4). the extended [OIII] emission is not comparable for radio galaxies and quasars  
\citep{Jackson}. Advocates of unified schemes 
interpret this to mean that some of the [OIII] emission is
obscured in the radio galaxies but not in the quasars, but this was
originally taken to provide evidence against the unified hypothesis.
Alternately, the [OII] emission, being lower ionization state gas, could be
ionized by other sources than the AGN (e.g., star formation); (5). other infrared 
studies of quasars and radio
galaxies in other redshift ranges \citep{Hes2} find more infrared emission from quasars
than radio galaxies. Thus, like the extended narrow line gas and the radio source size
tests, this test seems to give results that are dependent upon redshift and thus
inconclusive support for the unified schemes. So, while there is some significant support
for a unified scheme, there seems to be significant contrary evidence as well. And
because several of these tests yield
different results in different redshift ranges, an evolutionary scenario is suggested. 

In this work, we seek to test the evolutionary scenario for quasar fading by 
searching for indirect evidence of an
ICM forming around some of the radio galaxies and quasars studied in 
Paper 1 using distortions in their
extended radio structure. If the evolutionary hypothesis is correct, 
distorted radio structure  will be found
only for the lower $z$ cluster sources in our sample. We use this 
indirect methodology because very few
high-$z$ AGN-selected clusters have been detected directly by 
extended X-ray emission. Unambiguous detections
exist for only five sources in our sample \citep[e.g.,][]{GizLea,HW99,
WB00},  which are insufficient to draw firm conclusions about 
the intermediate redshift quasar and FR2
radio galaxy population as a whole.

Several previous studies have been conducted on the relationship 
between environment and FR type at low and
intermediate redshift. In an attempt to determine the relationship 
between FR1 and FR2 type radio galaxies,
\cite{OwenLaing} and \cite{OwenWhite}
examined very low redshift ($z < 0.2$) radio 
galaxies both in clusters \citep{OwenWhite} and
out of clusters
\citep{OwenLaing}.  They further subdivided the types of radio 
structure and found evidence for a third,
transitional (FR1/FR2) type, which  they refer to as ``fat doubles''. 
The lower power FR1s and the
transitional FR1/FR2s were found mostly in clusters while the higher power 
FR2s were found to avoid clusters. Similar
results concerning the effects of environment on radio structure were 
found earlier by \citet{Stoc78},
\citet{LS79}, \citet{Lilly87}, and \citet{Prestage}.   In work used by us in
Paper 1, HL extended these  studies of the 
environments of FR1 and FR2 radio galaxies to
$z \sim 0.5$.  Their results showed that higher power FR2 radio 
galaxies are found in richer galaxy
environments at higher
$z$ than at lower $z$, while the lower power FR1s show no change in 
environment with redshift. \citet{chap5}
confirm that FR1s  are found in rich clusters with a dense ICM 
regardless of redshift by observing with the
VLA a sample of X-ray selected  clusters out to $z\sim$0.8. While 
FR1s were found in distant clusters where a
dense ICM is demonstrably present,
Rector, Stocke \& Ellingson (1995) found evidence 
against the  presence of a dense  ICM around
a sample of quasars (FR2s) at these same redshifts. Because the 
\citet{Rector} work  used the same methodology
as employed herein, this paper will incorporate and expand upon  that 
work to include both quasars and radio
galaxies.


To investigate the evolution of cluster AGN further, in Paper 1 of 
this series we conducted an optical
imaging study for a large sample of 3CR/FR2 radio galaxies at 
0.15$\leq z \leq$0.65, quantifying cluster
richness for a very large percentage of the sample using the B$_{gg}$ 
formalism (see e.g., EYG). While most
FR2s imaged in Paper 1 have B$_{gg}<$500Mpc$^{1.77}$ (Abell richness 
class $<$0), a comparable percentage
($\sim$25\%) of all FR2s studied at $z$=0.15-0.65 are in richer 
environments (B$_{gg}$= 500-1200 Mpc$^{1.77}$
or Abell class 0-1). From that study we concluded that the cluster 
richness data offer significant evidence
against the \citet{Barthel}  hypothesis that quasars are ``beamed'' 
or less-obscured radio galaxies, since many FR2 radio
galaxies were found in clusters at epochs 0.15$\leq z \leq$0.45 
when no quasars are found in clusters.
Further, the cluster FR2s in this redshift range provide the 
``missing link''  between cluster quasars at
$z\sim$0.5 and cluster FR1s in the current epoch.  In Paper 1 we 
interpreted the result that  FR2s are found
in clusters in the past but not in the present epoch  as evidence for 
an evolving ICM in these AGN-selected
clusters. The fading of both the AGN optical brightness and the radio 
power (and its associated morphological
changes) are suggestive of evolution between the types of radio 
galaxies themselves. That is,  Paper 1
suggests that quasars at
$z\sim$0.5 fade to become FR2 radio galaxies at $z\sim$0.25 and then FR1s at 
$z\sim$0 on an ``e-fading'' timescale of 0.9
Gyrs (for H$_0$= 50 km s$^{-1}$ Mpc$^{-1}$).

In this paper we use the 3CR and quasar samples from Paper 1  and \citet{Rector} to 
investigate the extended radio morphologies of
these sources for confirmation of this ``evolutionary hypothesis''. 
That is, do we see changes in the radio
morphology of the 3CR/FR2 sample used in Paper 1 due to the changing 
ICM density expected in the ``fading
AGN'' hypothesis?  This paper is organized as follows: The AGN 
samples used in this study are discussed in
Section 2. The radio structure parameters used herein are defined and 
the morphology data are presented in
Section 3. These parameters are simple, quantitative measures of 
radio source size (and thus confinement)  and
distortions. The results of the correlation analyses between  the 
radio structure parameters and the density
of the surrounding galaxy environment ($B_{gg}$) are given in Section 
4.  This correlation analysis shows that
only some of the radio structure parameters are sensitive to the 
presence of a dense ICM. However, those
parameters that do correlate with B$_{gg}$ do so only at
$z\leq$0.4, strongly suggesting that the ICM around these sources is 
just forming at
$z\approx$0.4. Section 5 contains a  comparison between the clusters 
found around the FR2s in the present
sample  and rich, X-ray selected clusters at comparable redshifts, 
which contain only FR1 type radio sources
\citep{chap5}. Because the X-ray selected clusters are 
demonstrably richer than the AGN-selected
clusters in our sample, it is not surprising that both a dense ICM 
and also  FR1 radio galaxies are present in
them. A summary of the findings is given in Section 6, in which we 
report that this investigation provides
substantial, additional support for the ``evolutionary'' 
hypothesis of EYG and Paper 1.  Future work
that can further test this hypothesis   also is proposed.  Values of 
H$_0$ = 50 km s$^{-1}$ Mpc$^{-1}$ and
q$_0$ = 0 are assumed throughout this work to be consistent with 
earlier work;  however, none of the results
are sensitive to these choices.

\section{The 3CR and Quasar Samples}

The sample observed and analyzed for this study consists of all 3CR 
radio  galaxies and quasars with 0.15 $< z < $0.65 and
$|b_{II}| \ge 15^\circ$ that are listed in the Revised 3C 
Catalog of Radio Sources from Smith, Spinrad \& Smith (1976)
as updated by \citet{Spin85} and
\citet{Spinrad}. This sample is discussed in Paper 1 with the basic 
properties for sources in that sample
given in Table 1 of Paper 1. While the 3CR is a low-frequency radio-selected sample,
it is not rigorously complete due to the exclusion of a few very large 
or confused sources 
\citep[see][]{3CRR}.
But almost all of these potential additions are at lower $z$ and so do not affect the sample
investigated here.  A systematic exclusion of very large sources would 
cause a slight bias against discovering clusters, but given
the small number of possible additions to the 3CR, we expect this bias to
be negligible. There are 66 radio galaxies and 14 quasars listed in 
\citet{Smith}
in the redshift range of this study but one of the galaxies, 3C\,258, was 
discovered to contain what is believed
to be a distant background quasar in its spectrum (A. Dey, private 
communication).  Since the angular size  of 3C\,258 is
quite small for a source at a redshift of $z = 0.165$, it is likely 
that the radio source is associated with
the background object. Thus, this object has been removed from the 
sample.  All of the sources in this sample
have a rest frame luminosity  well above (26.8 $\leq log P_{178}$ (W 
Hz$^{-1}$)$\leq$ 29.0) the nominal
boundary between FR1 and FR2 type radio sources
\citep{FR, OwenLaing}. Further details of the sample selection 
and source properties  can be found in
Paper 1 and \citet{myPhD}.

Previously used as a comparison sample in Paper 1, the quasar sample of
\citet{YE93} contains 55 quasars with $z < 0.65$ and
$|b_{II}| \ge 30^\circ$ that are not found in the 3CR sample.
\citet{Rector} provide radio structure parameters for 24 of these 55 
quasars and we adopt their values
directly.  When combined with the 3CR sample, radio structure 
parameters are available for 65 radio galaxies
and 38 quasars; 63 of these radio galaxies and 35 of these quasars 
have a double-lobed radio structure (so all
their radio structure parameters are well-defined). Information on 
the environment (i.e., $B_{gg}$ values
taken from Paper 1 and Yee \& Ellingson 1993) exists for a very large 
subset of these sources: 50 radio
galaxies and 34 quasars. Because the few objects lacking $B_{gg}$ 
values were missed due to optical observing
conditions, no bias is introduced into this study due 
to their absence (see Paper 1 for
details).

\section{The Observational Data}

In order to measure accurately the radio structure parameters used in 
this study, it was necessary to obtain
radio maps showing the core, hot spots and extended structure of each 
source.  Often this required maps of the
same source at different frequencies because the steep-spectrum,  
extended structure is more easily detected at
L-band frequencies (1.3-1.7 GHz) while the flatter spectrum cores, 
jets and hot spots are better observed at
the higher frequencies of C (4.5-5.0 GHz), X (8.1-8.8 GHz) and U-band 
(14.6-15.3 GHz). Adequate maps of most
of the sources were found in the literature. However, we did observe 
15 of the radio galaxies used in this
study with the VLA in order to obtain new maps more suitable for our 
purposes \citep[see][]{4i}. 
These new data  include 20\,cm continuum maps to detect lower 
surface brightness extended structure as
well as a few higher frequency maps of smaller sources to locate 
previously undetected cores. Surprisingly,
one of the ``missing'' maps was a 20\,cm map of Hercules A (3C\,348), 
which can now be found in \citet{4i} 
and Section 4 herein.

The richness of the cluster environment is quantified using the 
amplitude of the galaxy-galaxy spatial
covariance function ($B_{gg}$) evaluated at the location of the radio 
galaxy. This quantity and its associated
errors are discussed in detail in Paper 1 and references therein. All 
$B_{gg}$ values used in this study are
taken from Paper 1 (for the 3CR sources) or from \citet{YE93} 
(for the non-3CR quasars).

\subsection{The Radio Structure Parameters}

To quantify the radio structure of each source, the following 
parameters were measured from the radio maps:
the largest angular size ($LAS$), the angular size ($\Psi$) measured 
between the brightest spot in each radio
lobe, the projected bending angle ($\beta$), the lobe length 
asymmetry ($Q$), and the Fanaroff-Riley ratio
($FR$ ratio). These quantities are defined in the text below and some 
are illustrated in Figure 1.  The radio
structural parameters for the 3CR sources used in this study are 
listed in Table 1; the non-3CR quasar data
can be found in Table 1 of
\citet{Rector}. Each of these parameters are defined and discussed in 
detail in the subsections just below.
Table 1 includes by column: (1) 3CR source number; (2) redshift; (3) 
largest angular size ($LAS$) in arcsecs;
(4) largest angular size ($LAS$) converted to  h$^{-1}_{50}$ kpc; (5) 
literature reference and band for the
map used in the $LAS$ measurement;  (6)  angular size between 
brightest flux points ($\Psi$)  in arcsecs; (7)
angular size between brightest flux points ($\Psi$) converted to 
h$^{-1}_{50}$ kpc;  (8) projected bending
angle ($\beta$) in degrees;  (9) literature reference and band for 
the map used in
$\Psi$ and $\beta$ measurements; (10) the lobe length asymmetry 
($Q$) with values ranging from 1.0 for
symmetric sources to $>>$1 for very asymmetric sources;  (11) 
Fanaroff-Riley ratio ($FR$ ratio) with values
ranging from
$\approx$1 (i.e., highly-collimated  FR2-like structure) to
$<$0.5, values typical of FR1s; (12) literature references and band 
for the maps used in $Q$ and $FR$ ratio
measurements; and (13) additional references for radio maps of these 
sources.  Because considerable effort was expended locating the best
available radio maps for this work, the listing of map references is
extensive.
Missing values indicate that the source does not
have a double-lobed structure (i.e., the source has either a core + 
jet type structure or is point-like).
Relativistic beaming and/or projection may affect some of the 
structures we see (e.g., one-sided jets, large
bending angles and source sizes). We do not take projection or 
beaming effects into account because we
believe that their effects are modest in our analyses (we show this 
explicitly in Section 4.1). This is both
because this sample was primarily selected at low frequency (178 MHz) 
and because we omit obvious examples of beamed sources (3C\,48,
3C\,93.1 3C\,196, 3C\,273 \& 3C\,345) from our analyses.

%

\subsubsection{Largest Angular Size} The $LAS$ (columns 3 \& 4 of 
Table 1) is defined by the largest straight
line separation between any two regions of the source showing 
detectable emission.  Although this quantity can
be dependent on instrumental sensitivity and map frequency, it is a 
better measure of the ``total'' size of the
source when the source contains emission that lies farther from the 
core than the lobe peaks.  Due to its
possible sensitivity and frequency dependence, $LAS$ was measured 
from all available maps of the same source
and the largest viable  measurement was recorded. The $LAS$ measurement
typically comes from an L-band map  and was
measured from the 3$\sigma$ map contour. For some of the 
smaller sources, however, the resolution of
the instrument (typically 1-3 arcsec for VLA L-band) artificially 
inflates this value (i.e., $LAS \sim$ beam
size) and in these cases a measurement taken from a higher resolution 
map was used, despite reduced
sensitivity to steep-spectrum emission. 
FR2 type structures have 
fairly robust measurements of $LAS$ because
the leading edges of each emission lobe are usually bright and are 
detected at high signal-to-noise ratio
(SNR).  However, FR1-type structures become dimmer further from the 
core and so their $LAS$ values may vary
substantially from map to map.   Since only a few of these sources 
have an intermediate FR2/FR1 radio
structure (see Section 4), most $LAS$ measurements do not vary much 
from map to map ($\pm$5\%). For some
sources, maps unaffected by undersampling were not available. Values 
taken from maps that may be undersampled
affect 16 sources only and are listed as lower limits in Table 1.

\subsubsection{Peak to Peak Angular Size} The angular size $\Psi$ 
(columns 6 \& 7 in Table 1)  is the angular
distance from the peak flux point in one lobe to the peak in the 
other lobe (see Figure 1) and is similar to
the $LAS$ for edge brightened sources.  Although not as good a 
measure of the ``total'' size of a source
(i.e., $\Psi \leq LAS$), this measurement is more robust than the 
$LAS$ measurement because it is relatively
independent of the sensitivity or frequency of the map from which it 
was taken.   The measurement is only
slightly affected by resolution and is not affected by undersampling. 
Values of $\Psi$ taken from different
maps of the same source typically have variations of only 2-3\%.  In 
general, $\Psi$ was measured from the
highest resolution maps available because these maps give the most 
precise position of the lobe peaks.

\subsubsection{Projected Bending Angle} The projected bending angle
$\beta$ (column 8 in Table 1) is defined to be the angle between the 
two intersecting lines that run from the
peak flux point of each lobe through the core (see Figure 1).  Beta 
is a measure of the non-collinearity of
the source.  This quantity was measured to the nearest degree. 
Measurements from different maps of the same
source give variations of 2-3$^{\circ}$.  This quantity is best 
measured from the highest resolution maps
available because these maps give the most precise peak and core 
positions and so these values were taken from
the same maps as $\Psi$. For 7 sources, no radio core was 
detected in any map, in which case the
optical position was used as the radio core position.  Since the 
errors in these optical positions are
typically
$\sim$ 1 arcsec, values calculated using optical positions are more 
uncertain than those using radio core
positions (which have typical errors of 0.1-0.3 arcsec). Therefore, 
all values calculated using an optical
position rather than a radio core position are marked with a colon. 
This uncertainty can be especially
important for the smaller sources ($\Psi \lesssim 10''$) where a 
small difference between the optical and
radio core position can result in a substantial change in the value 
of $\beta$ (and $Q$ as well; the $FR$
ratio is less sensitive to the core position).

\subsubsection{Lobe Length Asymmetry} The lobe length asymmetry $Q$ 
(column 10 in Table 1) is defined to be $Q
= h_1/h_2$ where
$h_1$ is the distance from the core to the farther lobe peak and
$h_2$ is the distance from the core to the closer lobe peak (see 
Figure 1). Measurements from different maps
of the same source give typical variations of 5-10\% in $Q$.  Because 
the values of $h_1$ and $h_2$ were also
used in the calculation of the $FR$ ratio (see below), they were 
generally taken from maps showing the largest
amount of extended structure. Occasionally, some of these maps had 
much poorer resolution than the maps from
which $\Psi$ and $\beta$ were measured. For these 17 cases, $Q$ was 
taken from the high resolution maps and
these values are marked with an asterisk in column 10 of Table 1. For 
most $Q$ values, the map reference and
frequency band are given in column 12. For $Q$ values marked with an 
asterisk, the map reference and frequency
band are given in column 9. Values of $Q$ marked with a colon were 
obtained using an optical position rather
than a radio core, as discussed above.

%
\subsubsection{Fanaroff-Riley Ratio} As discussed previously, the 
$FR$ ratio (column 11 of Table 1) was
developed by Fanaroff \& Riley (1974) to help classify the structure 
of radio galaxies.  It was originally
defined as $\Psi$/$LAS$ (see Figure 1).  However, the $FR$ ratio used in
\cite{Rector} and herein is slightly different. It is defined to be 
$FR = (h_1 + h_2) / (l_1 + l_2)$ where
$h_1$ and $h_2$ are defined as above and $l_1$ and
$l_2$ are the distances from the core to the farthest extent of the 
lobes (see Figure 1). Like the original
$FR$ ratio of Fanaroff \& Riley (1974), it is a measure of the 
position of the lobe peaks relative to the lobe
edges. For most sources, the two definitions of this ratio yield 
nearly identical values.  Differences can
occur for sources with large bending angles and for sources with an 
$LAS$ substantially larger than the
angular size measured along the radio axis (i.e., sources with large 
regions of extended emission transverse
to the radio axis). For some sources it was even unclear whether or 
not the farthest detectable emission
actually belonged to either of the lobes (e.g., 3C\,171). So, in 
order to remain consistent, we chose to
always make the $l$ measurements along the line containing the core 
and the lobe peak flux point. Values of
$FR$ computed with $l$ measurements substantially smaller  
($\lesssim 90\%$) than the distance to the farthest
detectable emission are marked with a dagger.

Because the $FR$ ratio can be dependent on the resolution, 
sensitivity and frequency of the map, values from
different maps  varied by 5-10\%.  In order to maximize sensitivity 
to the extended lobe emission,
measurements were usually taken from the L-band map. The few values 
taken from maps at higher frequencies
should be considered slightly  suspect because it is uncertain 
whether or not the full extended structure was
detected in these maps. Such values have no distinguishing mark (as 
the map frequency is given in column 12).
Measurements taken from maps that may be undersampled can 
underestimate the values of $l$ and so these values
of $FR$ are given as upper limits.  Values of $FR$ marked with a 
colon were obtained using an optical position
rather than a radio core, as discussed above.

%
%
%
%
\subsection{Comments on Individual Sources}

In general, comments on individual sources made herein  are 
restricted to those sources either with
difficulties and/or new, unpublished information which relates 
directly to  the measurement of the radio
structure parameters in Table 1. More detailed comments on these 
sources can be found in
\cite{myPhD}.  A few problems are generic to several sources and we 
discuss them first.  There are four sources in Table 1 (3C\,48,
3C\,196.1, 3C\,273 \& 3C\,345) with core-jet type structure and
one point source (3C\,93.1), so that the geometry
assumed by Figure 1 is not applicable.  Thus, several of the 
quantities described in Section 3.1 could not be
measured for these sources and so entries for them are left blank in 
Table 1.  As noted previously, several
sources have L-band maps which suffer from some undersampling; values 
taken from these maps are noted as
limits in Table 1.  Additionally, despite higher frequency maps being 
available, some sources have no detected
core. In this instance, values affected adversely are marked with 
colons. Unless the only available maps are
quite deficient and/or the source structure is quite complex, the 
errors introduced into the current
measurements are modest  (i.e., creating additional errors of only a 
few sigma). Only the most difficult cases
of the above problems are discussed individually below.

\noindent 3C\,28:  Although the optical position has been used for 
measurements, a very faint core has been
detected at 5 GHz (C band) that is coincident with the optical ID 
(J.T. Stocke, unpublished).






\noindent 3C\,93.1:  This source appears point-like in all maps 
available to us.  Thus, our value of $LAS$ is
given as an upper limit in Table 1  and values for the other 
parameters could not be measured, and so appear
as as blanks in Table 1.

\noindent {3C\,99}:  Because the L-band maps of this source have 
relatively modest dynamic range, the $LAS$ and size values reported
in Table 1 are quite uncertain.  Also, $Q$ varies from 2.3-4.7
depending upon the low-frequency map used (408 MHz or 1.4 GHz).
Therefore, the high $Q$ value for this source is very uncertain.

\noindent 3C\,225B: No core detection is known for this somewhat 
small source and so the optical position
marked on the C-band map of
\cite{4m} is used for all measurements. Since the angular size of 
this source is small (6.3$''$) and the
errors in the optical position are relatively large ($\sim 1''$), 
structural parameters measured using the
optical position are highly uncertain and the actual values could 
vary substantially from those quoted in
Table 1. Note, however, that none of the values listed in Table 1 are 
unusual.  The C-band map of \cite{4m}
shows two typical lobes.  The source is not resolved in the only 
L-band map available (Macdonald, Kenderdine
\& Neville 1968).

\noindent 3C\,275:  No core has been detected for this small source 
and so the optical position of \cite{2b}
was used for all measurements. This optical position differs from 
that of \cite{Spinrad} (which is listed in
Table 1 of Paper 1) by nearly 4$''$ in right ascension (the 
declinations agree).  The errors in the position
of \cite{2b} are given as 0.65$''$ while those of \cite{Spinrad} are 
given in Kristian, Sandage \& Katem
(1974) as $\sim$ 1$''$ (in each direction).  Since the positions do 
not agree within the errors, we have
chosen to use the position quoted to better accuracy.  Although the 
errors in this optical position are
smaller than most used in this work, the angular size of this source 
is so small (6.8$''$) that the structural
parameters measured using this optical position must be considered 
uncertain.  Note, however, that none of the
values listed are peculiar.  The C-band map of \cite{2k} shows two 
typical lobes.  No L-band map was available
for this source and we did not observe it in L-band because of its 
small angular size.



\noindent 3C\,288:  The southern lobe extends to the west away from 
the radio axis by more than the
peak-to-peak source size and then extends further to the south. The 
northern lobe continues beyond the peak to
the north by more than the peak-to-peak source size and also extends 
to the west (although not as far as the
southern lobe).  This northern extension gives the source its low 
$FR$ value.  Possible undersampling of the
map may hide even larger regions of extended emission. According to 
\cite{2g}, the radio core in their C-band
map is coincident with the optical position marked on their L-band 
map (the cross closer to the southern peak)
and so the $FR$ value taken from the L-band map is not marked with a 
colon.  The
$Q$ value measured from the L-band map is 32\% larger than the value 
taken from the C-band map listed in Table
1.  This large difference is probably due to differences in resolution.


\noindent 3C\,299: Due in part to the large lobe length asymmetry of 
this source, the eastern lobe of this
source was once thought to be a separate compact steep-spectrum 
source, which is why many maps show only the
eastern component.  Using the core position from the L-band map of
Leahy, Bridle \& Strom (1997),
the L-band map of \cite{3h} was used for $FR$ measurements 
because the L-band map of \cite{4a} is
probably undersampled.  The
$Q$ values from these two maps differ by 17\%, probably due to 
differences in resolution.  The $Q$ value from
the higher resolution map of
\cite{4a} is listed in Table 1.

\noindent 3C\,303.1: No core has been detected for this small source 
and so the optical position of
\cite{Spinrad} is used for all measurements.  Since the errors in the 
optical position are comparable to the
angular size of the source, structural parameters measured using the 
optical position are highly uncertain and
could vary substantially from those quoted in Table 1.  Note, 
however, that none of the listed values are
peculiar.  The X-band map of \cite{4h} is used for $Q$ and $FR$ 
ratio measurements because the L-band
maps show no additional extended structure.

\noindent 3C\,313: Maps of this giant source are almost certainly 
undersampled and it is likely that large
regions of emission are missing.  The values of $LAS$ and the $FR$ 
ratio are definitely suspect.  The eastern
lobe shows a small extension to the north but fully sampled maps are 
needed before the structure of this
source can be assessed accurately. The X-band map of \cite{3l} is used for
$Q$ and $FR$ ratio measurements because the L-band maps either show 
no additional structure or have poor
resolution.

\noindent 3C\,319: No core is detected for this source and so the
optical position marked on the L band map of \cite{r173} is used for
all measurements.  This optical position agrees with that of
\cite{Spinrad}.  Although the angular size of this source is large,
the values of $\beta$, $Q$ and $FR$ are all uncertain because, in
addition to the uncertain core position, the peak in the western lobe
is weak, not well-defined and lies relatively close to the optical
position.

\noindent 3C\,320:  All maps suffer from poor resolution and show 
only two featureless lobes.  The $LAS$ and
converted physical size taken from our map may be slightly 
overestimated due to both the poor resolution
and the elongated beam shape. A radio core is not apparent in any map 
but when the two lobes are removed from
our X-band map, a structure resembling a core remains (see Harvanek 
\& Hardcastle 1998 for further details).
An attempt was made to place this core position on the L-band map of 
\cite{2n} for $FR$ ratio measurements but
the coordinates of the L-band map appear to be inconsistent with 
those of our map because the core was not in
the same position relative to the lobe peaks.  Thus, for the $FR$ 
ratio measurements, the optical position
marked on the L-band map was used.  Values of $\beta$ and $Q$ taken 
from the L-band map agreed, within the
errors, with those listed in Table 1, which were taken from our 
X-band map.  This indicates that the optical
position marked on the L-band map of \cite{2n} is close to that of 
our radio core position even though the
coordinates are discrepant.  The low $FR$ ratio (0.46) may be due to 
the poor map resolution rather than the actual radio structure and
so is listed as uncertain (marked with a ``?'') in Table 1.

\noindent 3C\,459: The $Q$ value measured from our L-band map is 25\% 
larger than the value listed in Table 1,
which was taken from the U-band map of van Breugel (private 
communication). However, the C-band map of
\cite{3e} detects a weak western component, not seen by any other 
observations. If this component is real,
then $Q\leq$1.5. Therefore, the large $Q$ value for this source is 
very uncertain.

\noindent 3C\,460: The difference in the $Q$ values measured from the 
U- and L-band maps of van Breugel
(private communication) is substantial (the U-band map value is 32\% 
larger).  Most of this discrepancy is
probably due to the fact that the resolution of the U-band map is $10
\times$ better (which is why the U-band value is listed in Table 1). 
Since no core was detected in the L-band
map, measurements from this map were made using the core position 
from the U-band map and so differences in
the coordinates of the two maps also could contribute to the 
discrepancy in the $Q$ values.
The low value of the $FR$ ratio may be due, at least 
partially, to the L-band map resolution.

\section{Correlations With the Radio Structure Properties}

\subsection{Comparison Between Radio Galaxies and Quasars}

In Paper 1 we compared the galaxy environments of radio 
galaxies and quasars in order to test the two
hypotheses that relate these two types of objects.  We found that the 
orientation angle hypothesis of
\cite{Barthel} cannot account for the difference in the clustering 
environments of the two types of objects at
$z < 0.4$ seen in
Paper 1, whereas the evolutionary hypothesis of EYG
actually predicts the difference seen.
This result appears to rule out Barthel's orientation angle 
hypothesis as the primary means of relating
quasars and radio galaxies, and favors the ``evolutionary 
hypothesis'' of EYG.  However, orientation angle may
still play a secondary role in this relationship (i.e., SOME quasars 
are beamed radio galaxies, like the
core-jet sources mentioned above).

While the radio galaxy and quasar samples used here  are not matched 
precisely in redshift and radio power,
we  compare their radio structures here for two reasons: (1) to 
verify that  orientation effects do not
dominate the results found (in which case the differences in size and 
morphology between quasars and radio
galaxies would be large) and (2) to make certain that  there are no 
systematic biases in  measurements of
\cite{Rector} for the quasars and our own 3CR radio galaxy and quasar 
measurements in Table 1.

A comparison of the quasar and radio galaxy distributions for each of 
the radio structure properties in Table
1 was done using a Kolmogorov-Smirnov test (KS test hereafter). The 
likelihood that the quasar and radio
galaxy distributions came from the same parent population is 35\% for 
the projected physical size, 48\% for
the projected bending angle, 98\% for the $FR$ ratio and 92\% for the 
lobe length asymmetry.  
The latter two properties ($FR$ and $Q$) are not expected to show 
an orientation
angle effect, but neither is there any 
significant evidence for quasars and radio
galaxies being systematically different in the other two properties; 
e.g., the difference in projected
physical size distributions between our two samples is $\sim$ three 
times less significant than the difference
found by Barthel (1989) for his samples. We do {\it not} mean to challenge
Barthel's result by our analysis because we have not attempted to match our quasar and
radio galaxy samples as closely as is necessary to conduct this test 
with accuracy, and there already are published conflicting results in the
literature (see references cited in the Introduction). 
For our quasar and radio galaxy samples,  
both the mean and median of the the projected physical
size and bending angle distributions were compared and 
the differences were small ($ < 1
\sigma$). Thus, we feel justified in combining these two samples into 
a single ``FR2'' sample.  From power
level considerations, all of these quasar and radio galaxy sources 
are FR2s, and so the combined sample is
ideal to search for evidence of an ICM using distortions in the radio 
structure, because an FR1-like radio
structure (e.g., low $FR$ ratio) for a source in this sample cannot 
be attributed to a low radio power (see
Table 1 in Paper 1 for P$_{178}$ values).

\subsection{Radio Structure vs. $B_{gg}$}

The search for evidence of an ICM around these sources is performed 
by looking for correlations between the
richness of the galaxy environment ($B_{gg}$) and the radio structure 
properties that could be affected by
interactions with a dense surrounding medium.  This technique 
implicitly assumes that a richer galaxy
environment possesses a denser ICM,  which may not be the case at all 
redshifts included in the sample. A
dense ICM interacting with the radio structure should tend to 
confine and distort the structure, making it smaller, more bent
(larger $\beta$) and more asymmetric (larger $Q$) and thus would lower
its physical size and FR ratio.

The overall comparisons using the combined quasar + radio galaxy 
``FR2'' sample shows little correlation
between the environment and the radio structure parameters, yielding 
correlation coefficients of $r = -0.01,
-0.01, -0.15$ and
$-0.13$ for $B_{gg}$ with the physical size, $\beta$,
$FR$ ratio and $Q$, respectively (with confidence levels of $ < 50\%, < 50\%,
\sim 85\%$ and $\sim 80\%$, respectively).  Since the weak 
correlation between environment and $Q$ is in the
opposite direction expected for an interaction with a dense medium 
(and is due primarily to three sources with
uncertain $Q$ values; see Section 3.2), the only evidence at all for 
a dense ICM using these statistics is the
weak anti-correlation between environment and FR ratio.  Therefore, 
we conclude either that the morphology of
FR2s is unaffected by the surrounding ICM  or that a dense ICM has 
yet to form around those sources with high
$B_{gg}$ values in our sample.

However, it is possible that the ICM is actually forming in these 
clusters during the epochs studied here
(i.e., $0.15 < z < 0.65$; see Section 1) and so its effects may be 
detectable only after a certain epoch.
Indeed, this is the prediction of the ``evolutionary hypothesis'' 
discussed in the Introduction; i.e., when
the ICM begins to develop, the quasars fade, leaving only radio 
galaxies in clusters at the lower redshifts.
The further fading of FR2s into FR1s would then be accompanied by 
observable morphology changes at and below
the  redshifts where the ICM begins to form. Given the richness of 
the clusters (Abell class 0-1) we have
found around these FR2s,
$z$=0.4 is a reasonable epoch in a low-$\Omega_{matter}$ universe 
for the formation of a dense ICM (Perrenod 1978; Stocke \& Perrenod
1981; Eke, Cole \& Frenk 1996).

In order to investigate this possibility, the data were divided into 
low ($z < 0.4$) and high ($z > 0.4$)
redshift subsamples and the correlation analysis was performed for 
each subsample separately.  The separation
at
$z = 0.4$ was chosen both because it nearly divides the sample evenly 
and because of the results of Paper 1 and EYG; i.e., at
$z\leq$0.4 only radio galaxies, not quasars, are found in clusters. 
Figures 2, 3, 4 \& 5 show plots of the
various radio structure parameters vs. environment for both the low- 
and high-z subsamples. The correlation
coefficients for the data in each plot are given in the figure captions.

The plots of projected physical size ($\Psi$ converted to physical 
dimensions)  vs. environment ($B_{gg}$)  in
Figure 2 show some evidence for an ICM, but only at the lower 
redshifts.  The low-z panel shows a correlation
although both the coefficient and the confidence level are somewhat 
low ($r = -0.16$; confidence level of
$\sim 68\%$).  The correlation is such that smaller sources tend to 
be found in richer environments and, taken
alone, is slight evidence for an ICM around the higher
$B_{gg}$ sources at $z < 0.4$.   The high-z panel shows no correlation.

%

The presence of a dense ICM is most noticeable in Figure 3, which plots
$FR$ ratio vs. $B_{gg}$. While very few FR2s at $z > 0.4$  have $FR$ 
ratios that approach the 0.5 dividing
line between FR1 and FR2 morphology, over 25\% of the low-z subsample 
have $FR$ ratios near the dividing line,
including almost all the sources with $B_{gg} >$ 500 Mpc$^{1.77}$. 
The low-$z$ panel of Figure 3 shows a
relatively strong correlation between
$FR$ ratio and environment ($r = -0.39$; confidence level of $\sim 
99\%$). The correlation is such that
sources with a lower $FR$ ratio are found almost exclusively in 
richer environments and is strong evidence for
an ICM  around the sources with high $B_{gg}$ values at
$z < 0.4$.   The high-z panel shows no such correlation. Therefore, 
the weak correlation seen in this
relationship over the entire redshift range is due entirely to the $z 
< 0.4$ subsample.

%


The other two parameters, lobe length asymmetry and bending angle, 
show no correlation with environment at any
redshift covered by this sample. While the low-$z$ panel of Figure 4 
appears to show a slight anti-correlation
between lobe length asymmetry ($Q$) and environment, the correlation 
coefficient is very weak ($r = -0.09$)
and the confidence level of $ < 50\%$ suggests it is likely not 
significant;  the high-$z$ panel indicates a
similar strength anti-correlation.  In both cases it is obvious 
that most of these correlations come from
points representing uncertain data; i.e., the single high $Q$ point 
(3C\,99) at high-$z$ and two high $Q$ points (3C\,459 and 3C\,460)
at low-$z$ (see Section 3.2). Without these points there is no
correlation in either redshift range.  Also, these anti-correlations
are in the opposite sense to the expected effect that a
denser ICM would have on the lobe length asymmetry.

The low-$z$ panel of Figure 5 appears to show an anti-correlation 
between bending angle ($\beta$) and
environment but the coefficient indicates that it is not significant 
($r = -0.01$ at a confidence level of $ < 50\%$). The high-$z$
$\beta$ vs. B$_{gg}$ plot shows a similarly weak positive 
correlation.  Three of the four high-$\beta$ points
($\beta \ge 40 ^\circ$) in Figure 5 are very distorted sources with
poorly determined $\beta$ values \citep{Rector}.  The one
high $\beta$, high B$_{gg}$ point is the highly-distorted quasar 
3CR\,215.   Thus, neither $Q$ nor
$\beta$ show any evidence for being affected by an ICM surrounding 
some of these sources.

%

The results of the correlation analyses for $\Psi$ and $FR$ vs. 
$B_{gg}$ are evidence for an ICM  at $z
\lesssim 0.4$, but not at
$z \gtrsim 0.4$,  despite the high $B_{gg}$ values for many clusters 
around quasars and radio galaxies in that
redshift range.   The lack of the appropriate
$\beta$-$B_{gg}$ and $Q$-$B_{gg}$ correlations in any redshift 
interval covered by our samples are indications
that large
$\beta$ and $Q$  are caused by something other than an interaction 
with a dense ICM. For example, such
structural features may be due to a redirection of the radio jet 
and/or lobe caused by an inelastic collision
between the radio jet/lobe and a nearby galaxy or dense intergalactic 
cloud as discussed in Stocke et~al.\ (1985). That paper
predicted that inelastic jet-cloud collisions can cause both large
$\beta$ and $Q$.

The lack of a correlation between $\beta$ and $B_{gg}$  is also 
important because both previously (Hintzen \& Scott 1978)
and currently (Blanton et~al.\ 2000),
a large $\beta$ has been proposed as an indicator for 
rich clusters around powerful radio
sources. If this were true, bent radio sources would be a very 
powerful method for discovering extremely
distant clusters of galaxies. As Figure 5 shows, sources with 
$\beta > 20 ^\circ$ occur at widely different
$B_{gg}$ values, ranging from
$B_{gg} < 0$ to $B_{gg} > 1000$ Mpc$^{1.77}$.  Thus, an FR2 type 
radio source with a large bending angle does
not appear to constrain the richness of its  environment at all. We 
conclude that, while the ``bent''
morphology of FR1 type sources is strong evidence for a dense ICM, a 
large $\beta$  in an FR2 source is not an
unambiguous indicator of a rich cluster environment.

The \citet{Blanton} results utilize the radio survey called ``Faint 
Images of the Radio Sky at Twenty
centimeters'' (FIRST), which has a nominal survey limit of
$\sim$ 1 mJy. Since detection of a core and two lobes are required to 
define a large $\beta$, a minimum total
flux of 5-10 mJy is required for this  determination in FIRST data. 
Because the most luminous FR1s have log
P$_{20}~\approx$ 25.3 W Hz$^{-1}$, the maximum redshift at which FIRST 
can detect an FR1 is $z \approx$ 0.5. Beyond
that redshift, any bent source in FIRST would be an FR2, not an FR1. 
Based upon the analysis presented herein,
which fails to find any correlation between large
$\beta$  and high $B_{gg}$ value for FR2s, a deeper survey than FIRST 
would need to be performed in order to
discover rich clusters at $z \geq$ 0.5 using this method.

To illustrate the morphological differences indicated by the 
correlation between
$FR$ ratio and environment at low redshift ($0.15 < z < 0.4$), radio 
maps of four 3CR sources in rich
environments ($B_{gg} >$500 Mpc$^{1.77}$) and four 3CR sources   in 
poor environments ($B_{gg}
\sim 0$ Mpc$^{1.77}$) in this redshift range are shown in Figures 6 
and 7, respectively. Based on their radio
power, all sources in Figures 6 and 7 are easily within the range of 
FR2 type sources.  Specifically, the four
sources in rich environments have the following logarithmic radio 
power levels at 178 MHz in W Hz$^{-1}$ in
parentheses: 3C\,173.1 (27.90); 3C\,348 (28.66); 3C\,401 (27.66) and 
3C\,346 (27.16).   Note that 3 of the 4
sources in rich environments (Figure 6) have a low $FR$ ratio ($\sim 
0.5$) which is characteristic of FR1 type
radio structure.  The brightness and prominence of the jet(s), the 
absence of very well-defined, leading-edge
hot spots and the overall breadth and distortion of the lobes  of 
these 3 sources are also similar to FR1 type
structure, despite their FR2 power levels. Thus, these three sources 
are examples of the morphological type
originally described by Owen \& Laing (1989) as ``fat doubles'' or 
FR1/FR2 hybrid sources. In Figure 6, only
the source in the upper left panel (3C\,173.1) lacks some obvious FR1 
morphological characteristics. In
contrast, the 4 sources in poor environments (Figure 7) all have an $FR$ ratio
$\sim 1$ (i.e., typical of FR2s) and lack other FR1 type 
characteristics as well. Thus, it is apparent that at
redshifts of $0.15 < z < 0.4$, sources with FR2 power levels are much 
more likely to have some FR1 type
structural characteristics if they are in rich galaxy environments. 
Not only is this evidence of an ICM in the
richer environments at $z < 0.4$, but their ``fat double'', or
intermediate FR1/FR2, morphologies suggests that
these FR2 radio sources may eventually evolve into  FR1 sources at 
later times, as the cluster ICM thickens
around them.

Figure 6 shows only 4 of the 6 sources with $0.15 < z < 0.4$ and 
$B_{gg} > 500$ Mpc$^{1.77}$.  The other 2
sources also have FR1 type characteristics.  3C\,28 has a somewhat high
$FR$ ratio (0.73) but has a two-sided jet and extremely ``fat'' and 
distorted radio lobes (both of which are
characteristics of FR1s). 3C\,320 has a low $FR$ ratio (0.46) and 
might be described as a ``fat double'', but
this measurement and impression may be due to poor map resolution. 
Of these 6 sources, 3C\,173.1 (Figure 6,
upper left panel) is the only source lacking some  FR1 type 
characteristics and so the maps in Figure 6 are a
fair representation of sources with high $B_{gg}$ at
$z\leq$0.4.

Although not all sources with $0.15 < z < 0.4$ and $B_{gg}
\sim 0$ Mpc$^{1.77}$ have $FR \sim 1$ and lack FR1 type 
characteristics, most do and so the maps of Figure 7
are also representative of the low-$B_{gg}$ subsample.  Also, all 
maps in Figure 7 except 3C\,18 (upper left panel)
are quite colinear;  3C\,18 has a bent or ``dog-leg'' type structure 
with specific features attributable  to a
jet-cloud collision \citep{SBC}. For example, the lobe closer to 
the core is significantly broader than the
other lobe, as expected from an inelastic collision.  This source is 
a good example of why a large $\beta$ is
not an unambiguous indicator of a rich (or poor) cluster around an FR2 
type radio source (Figure 5).  Although
this source clearly has a large bending angle ($\beta=25 ^\circ$) the 
surrounding galaxy environment is quite
poor ($B_{gg}=-120$ Mpc$^{1.77}$; i.e., its surrounding galaxy 
surface density is actually less than the
average background for its redshift).
As mentioned above, at least four of the cluster radio sources at
$z\leq$0.4 in our sample (the three in Figure 6 plus 3C\,28) can be 
described as ``fat doubles''
(Owen \& Laing 1989), and one additional cluster FR2 source (3C\,320)
could be ``fat'' but the map resolution is
insufficient to be certain.  Additionally, three other FR2s at $z 
<$0.4 with an $FR$ ratio $<$ 0.5 can be
described as ``fat doubles'' (3C\,213.1, 3C\,219 and 3C\,288) but lack 
deep imaging to determine if a cluster
is present. We predict that: (1) a higher resolution, L-band map of 
3C\,320 will find that it has similar
morphology to the sources in Figure 6 and (2)  clusters will be 
discovered around 3C\,213.1, 3C\,219 and
3C\,288. Conversely, only 3C\,173.1 is found in a rich galaxy 
environment and is not ``fat'' and, no
non-cluster 3CR source at $z\leq$0.4 is demonstrably ``fat''. The 
only quasar in our sample which might be
described as a ``fat double'' is 3C\,215 ($B_{gg}$=1000 Mpc$^{1.77}$ 
at z=0.411). Recent ROSAT and CHANDRA
X-ray observations confirm the presence of a dense ICM around 4 of 
the 7 ``fat doubles'' at $z\leq$0.4:  3C\,28
\citep[][using an Einstein image]{3b}, 3C\,348 \citep{GizLea}, 3C\,219
\citep{HW99} and 3C\,346 \citep{WB00}.  Other claimed ICM detections 
were made with the ROSAT HRI around
$z\geq$0.5 FR2s (Hall et~al.\ 1995; Wan \& Daly 1996; Hall, Ellingson
\& Green 1997; Hardcastle \& Worrall 1999) with measured X-ray 
source sizes close to the resolution limits
of the HRI (particularly once spacecraft ``wobble'' during long 
exposures is taken into account; see Morse 1994 and Rector, Stocke
\& Perlman 1999).
Thus, we judge all these detections to be 
tentative. Nevertheless, recent CHANDRA
observations of    3C\,220.1 ($z$=0.620;
$B_{gg}$=451 Mpc$^{1.77}$; $FR$ ratio$\leq$0.88) obtained by
\citet{WBH} and 3C\,295 ($z$=0.461; $B_{gg}$=1030 Mpc$^{1.77}$; $FR$ 
ratio=0.78) obtained by \citet{Harris}
have detected definite extended cluster emission.  The 3C\,295 
detection is expected from its high-$B_{gg}$
value and from its radio structural parameters in Table 1.
However, the 3C\,220.1 detection is unexpected given its
rather modest
$B_{gg}$ value and it classical double radio structure at high-$z$. 
While it is possible that this FR2 is
projected onto a cluster \citep[i.e.,  3C\,220.1 is in the outskirts of a 
cluster similar to the case of Cygnus A;][]
{Owen97}, the discovery of even one or two more cases like 
3C\,220.1 (i.e.,  high $z$, classical FR2
structure but with a dense ICM) would be significant evidence against 
the inferences about the presence or
absence of a dense ICM made herein. With that in mind,  imaging 
spectroscopy with CHANDRA should be obtained
for the other few 3CR FR2s with tentative ROSAT detections 
\citep[see][]{HW99}. If the inferences herein are
correct, no dense ICM should be detected around 3C\,334 and 3C\,275.1 
(although this latter case is a less obvious prediction because
there is a significant galaxy cluster with $B_{gg}$=1125 Mpc$^{1.77}$ 
but the source is not demonstrably
``fat'') but extended X-ray  emission should be found surrounding 
3C\,215 ($z$=0.411; $B_{gg}$=1000
Mpc$^{1.77}$;
$FR$ ratio=0.65).


\subsection{Are the Observed Correlations Due to Radio Power or Redshift?}

Because the vast majority of the sources studied 
herein come from the flux-limited 3CR
Catalog, there is a strong correlation between radio power and 
redshift in our sample.  Since Paper 1
and EYG found correlations between
$B_{gg}$ and redshift for large subsets of our sample,  the 
morphology-$B_{gg}$ correlations found in the
previous section could be due to radio power instead.  So, in this 
section, we investigate the possibility
of correlations between the radio structure properties and both the 
radio power and the redshift.  The
correlation analyses were performed over the entire redshift range 
($0.15 < z < 0.65$) and also over both
redshift subsamples ($z < 0.4$ and $z > 0.4$) separately.   Dividing 
the data into two redshift subsamples
provides a comparison of the evolution over these two epochs which 
may be different if the ICM around these AGN is indeed
forming at
$z \sim 0.4$ as suggested above.

The analyses relating the various radio structure properties to the 
radio power revealed no significant
correlations. We find only a very weak correlation with 
r=+0.127 at a confidence level of 77\%
between radio power and
$FR$ ratio for our full sample. The correlations were even poorer 
for the  high and low-$z$ subsamples separately
(r=+0.047 and -0.057 respectively; both at a $<$50\% confidence 
level).  Since the $FR$ ratio vs.
$B_{gg}$ anti-correlation is in the low-$z$ subsample, the above 
correlations (particularly the last one)  are
far too weak to explain the correlation found in the last section. 
Also, we find no significant correlation
between radio power and projected physical size or $Q$, either for 
the full sample or for either subsample.
Bending angle is weakly anti-correlated with radio power in the full 
sample (r=-0.112 at 71\% confidence
level), but even more weakly correlated in the subsamples. While 
correlations between radio structure and
radio power level would be expected from the  classification scheme of
\citet{FR}, all the sources studied here are well above the nominal 
FR1/FR2 dividing line.  Thus, since no
truly low power sources (FR1s) are present in our sample, it appears 
that the range of radio powers is not
large enough to detect these previously known correlations. However, 
these results do show that a variation in
radio power over the range represented by the sources in this study 
($26.8 \le$ log
$P_{178}$ (W Hz$^{-1}$) $\le$ 29.0) is not sufficient to 
significantly influence the radio structure.  This
indicates that variations in radio power are not the dominant factor 
in causing the distortions in the radio
morphology.

The relationships between the various radio structure properties and 
redshift were also examined.  A
correlation between $FR$ ratio and redshift was found for the low-$z$ 
data and all data, both having a
coefficient of $r = 0.27$ at confidence levels of $\sim$ 95\% and 
99.5\%, respectively.   No correlation was
detected for the high-$z$ data alone.  This result shows 
that there is no evolution in this
property until $z \sim 0.4$, at which point the $FR$ ratio decreases 
with decreasing $z$, consistent with our
earlier interpretation that a dense ICM is forming  around these sources at $z
\sim 0.4$. No other correlations between radio structure properties 
and redshift are found except for a
marginal correlation between projected physical size and redshift ($r 
= -0.16$ at a confidence level of $\sim$
68\%) in the high-$z$ subsample alone. If real, this marginal 
anti-correlation could be due to radio sources
expanding with time until
$z\sim$0.4, when the developing ICM begins to confine them more 
strongly.  The lack of correlations between
$\beta$ and redshift and
$Q$ and redshift are a further indication that the ICM is not the 
dominant factor in creating large $\beta$
and $Q$.

We conclude that there is no strong evidence that radio power is 
responsible for the morphological differences
seen at
$z\leq$0.4 in our sample.  While overall power level, and thus jet 
power level, might be expected to make
morphological differences in extended radio sources, this cannot be 
the cause of the differences found here.

\section{Comparisons with X-ray Luminous Clusters}


A confusing factor in the interpretation of these results is that FR1 
type sources have been discovered in
clusters at redshifts comparable to the range studied here \citep[e.g., 
HL and][]{chap5}. In this
section we show that the clusters which contain FR1s at $z$=0.15 to 
0.83 are demonstrably richer than those
found around the FR2 quasars and radio galaxies in Paper 1.
Since the clusters  found to contain FR1s by
\citet{chap5} are all X-ray luminous, there is direct evidence that 
these clusters already contain a dense ICM.

The {\it Einstein} Extended Medium Sensitivity Survey
(hereafter EMSS; Gioia et~al.\ 1990; Stocke et~al.\ 1991;
Maccacaro et~al.\ 1994)
is one of the largest and best studied surveys for 
faint X-ray sources made to date. The EMSS
surveyed nearly 800 deg$^{-2}$ of sky, discovering 835 faint sources 
at 5 x 10$^{-14} <$ f$_x$ (ergs cm$^{-2}$
s$^{-1}$) $<$ 3 x 10$^{-12}$. Over one hundred of these sources are 
identified as distant clusters of
galaxies, including some objects at substantial soft X-ray 
luminosities ($>$10$^{45}$ ergs s$^{-1}$) and
redshifts ($>$0.8).

The EMSS cluster sample observed by \citet{chap5} with the VLA is 
composed of 19 EMSS clusters in the redshift
range $0.30 < z < 0.83$. Since this sample is small and does not 
match the redshift range studied herein, an
expanded sample, which includes all EMSS clusters with
$z > 0.15$ (59 sources), is also discussed.
\citet{chap5} provide detailed information (maps, fluxes, optical 
IDs, etc.) on the radio sources detected in
their $z >$0.3 EMSS cluster survey.  All sources known to be 
associated with these EMSS clusters have an FR1
type radio structure and radio power level.
For the 40 additional clusters in 
the extended sample of all EMSS clusters
with $z > 0.15$, examinations by one of us (JTS)  of the VLA C-array 
snapshot data summarized in
\citet{GL94} found only two sources with radio power near that of 
an FR2 (log P$_{178} \gtrsim 26$ W
Hz$^{-1}$). Both of these sources were unresolved in the 
snapshot data and with optical counterparts
as yet unconfirmed as clusters members, so that further observations 
must be made before the presence of FR2
radio sources in these EMSS clusters can be ruled out completely. 
Therefore, it appears very likely that all radio
sources in EMSS clusters at $z > 0.15$ have an FR1 type radio 
structure and power level.

The $B_{gg}$ values for the entire EMSS cluster sample, along with 
cluster names and redshifts, are listed in
Table 2.  Clusters marked with an asterisk are those in the sample of 
\citet{chap5}. If available,
$B_{gg}$ values provided by \citet{YE02} are used. These values were 
computed in the same manner as the
$B_{gg}$ values of Paper 1 and
\cite{YE93} and are identified by a ``Yee'' in the ``Comments'' 
column.  Otherwise, we have estimated the
$B_{gg}$ value from the X-ray luminosity as follows: the cluster 
X-ray luminosity was converted to a central
galaxy density,
$N_{0.5}$ (the number of excess galaxies with
$M_V \leq$ -19 within 0.5 Mpc of the cluster center), using the 
correlations ($r = 0.74$) given in
\citet{AK83} and \citet{YE02}.   The $B_{gg}$ values were then 
obtained using the
$B_{gg} = 32 N_{0.5}$ scaling  needed to convert $N_{0.5}$ values into the
$B_{gg}$ values supplied by \citet{YE02}.  
This is consistent with the scaling
factors derived in Paper 1 and in HL.
Cooling flow clusters are denoted by a ``CF'' in the 
``Comments'' column.  Because a cooling
flow appears to add ``extra'' X-ray emission to a cluster (increasing 
the X-ray luminosity per unit central
galaxy density), a downward (factor of two) correction was made to the
$B_{gg}$ value of the one cooling flow cluster (MS0735+74) with no measured
$B_{gg}$ value  by \citet{YE02}.  

In Figure 8, the $B_{gg}$ distributions of the two EMSS samples (left two
histograms) are shown compared to AGN-selected clusters in this survey
(right two histograms) with the $z\geq$0.3 only samples at the top and the
full samples at bottom.
Although the distributions of
AGN-selected and X-ray-selected cluster richnesses overlap in the range
$B_{gg}$=500-1000 Mpc$^{1.77}$, the EMSS clusters in the mean are 
much richer than the environments of the
sources in the 3CR radio galaxy and quasar samples.
%
%
KS tests were performed to quantify the differences in these
distributions.
For the $z >$0.3 samples, KS-test probabilities were found with 
only  2$\times$10$^{-5}$\%,
2$\times$10$^{-3}$\% and 8$\times$10$^{-5}$\% likelihoods that the 
EMSS distribution came from the same parent
population as (respectively)  the full radio galaxy$+$quasar sample, 
the radio galaxies alone, and the quasars alone.
For the full $z>$0.15 EMSS
\& AGN samples, these KS probabilities are all $< 10^{-7}$\%.

So, EMSS clusters, which are moderately rich to extremely rich 
environments ($B_{gg} \sim 500 - 2700$
Mpc$^{1.77}$), and which are known to possess a dense ICM, contain only FR1 type 
radio sources.  At the same redshifts,
the FR2 type radio galaxies and quasars are found in poor to 
moderately rich environments ($B_{gg} < 0$ to
$B_{gg} \sim 1200$ Mpc$^{1.77}$). However, some of the radio galaxies 
found in the richer environments
($B_{gg} > 500$) at $z \lesssim 0.4$   appear to have an FR1/FR2 
transition type structure (see Figure 6),
which we argue is indirect evidence of a dense ICM forming in these 
environments.  We expect, but cannot yet
prove, that the clusters found around FR2s at
$z>$0.4, do not yet contain a dense ICM. Thus, all current radio 
galaxy data are consistent both with FR1s
being exclusively present when a dense ICM is present and also with 
FR2s being present when a dense ICM is
absent. ``Fat Doubles'' (intermediate FR1/FR2s) are present when a 
dense ICM has first formed.
Since X-ray selected rich 
clusters have now been discovered out to
$z\sim$1 (Rosati et~al.\ 1998), the formation of a dense ICM in poorer 
clusters at $z\sim$0.4 is consistent with  the
theoretical expectation that the epoch of ICM formation is dependent 
upon the richness of the galaxy environment (e.g., Perrenod 1978).

\section{Conclusion}
\subsection{A Summary of the Findings}

In Paper 1 of this series, evidence was presented that quasars and 
FR2 radio galaxies are primarily related,
not by orientation as previously proposed
by \cite{Barthel}, but by evolution (EYG); i.e., quasars in clusters at
$z\sim$0.5 fade to become FR2 radio galaxies at $z\sim$0.25, and 
continue to fade to become FR1 radio galaxies
at
$z\sim$0. In order to further test the ``evolutionary hypothesis'', 
we have scrutinized the radio morphologies
for a large sample of quasars and radio galaxies at $z$=0.15-0.65 to 
search for indirect evidence of a dense
ICM around some of these sources.  This was accomplished using various 
morphology parameters that measure the
confinement and distortion of the extended radio source. We then 
determined whether these morphology
parameters were correlated with a quantitative measurement of galaxy 
richness ($B_{gg}$) from Paper 1. These
morphology  parameters included: projected  physical size, hot spot 
location (Fanaroff-Riley or $FR$ ratio),
projected bending angle ($\beta$) and lobe length asymmetry ($Q$). 
Using the radio structure measurements and
$B_{gg}$ values for a combined sample of 63 radio galaxies and 35 
quasars in the above redshift range we find
the following results:


1.  When the radio structure data for the radio galaxies and quasars 
are divided into low-$z$ ($<$0.4) and
high-$z$ ($>$0.4) subsamples, correlations between environment and 
projected physical size and between
environment and $FR$ ratio show evidence for the presence of a dense 
ICM in the richer environments ($B_{gg} >
500$ Mpc$^{1.77}$) only at $z
\lesssim 0.4$.  There are no correlations and, thus no evidence for 
an ICM around the sources at $z \gtrsim
0.4$. This suggests that the formation of a dense ICM in environments of
$B_{gg} \sim 500-1000$ Mpc$^{1.77}$ occurs at $z \sim 0.4$.

2.  There is no correlation between projected bending angle and 
environment or between lobe length asymmetry
and environment at any redshift within the sample.  This indicates 
that bending angle distortions and
asymmetry distortions in the radio structure are not caused by an 
interaction with the ICM. Instead, a
collision between the radio jet/lobe and a nearby galaxy or dense 
intergalactic cloud may be responsible
\citep[e.g.,][] {SBC}.

3. The lack of a correlation between bending angle and environment 
also demonstrates that for FR2 type radio
sources, a large bending angle is not a reliable predictor of a rich 
galaxy environment.  Thus, it should not
be used as a indicator of a cluster of galaxies 
around an FR2 at high-$z$, as has
been previously proposed by
\citet{Hintzen} and recently reproposed by \citet{Blanton}.


4.  No significant correlations between radio structural properties 
and radio power are found within the
range of radio powers used in this study: 26.8
$\le$ log P$_{178}$ (W Hz$^{-1}$) $\le$ 29.0.   Specifically,  all 
sources studied are FR2s, significantly
above the FR2/FR1 dividing line in radio power. Thus, the differences 
in radio structure seen in our sample
are not due to variations in radio power  but rather due to 
interaction with a dense ICM.

5. At $z <$0.4, the sources found in clusters with
$B_{gg}\geq$500Mpc$^{1.77}$ almost exclusively have morphologies that 
can be described as ``fat doubles''
\citep{OwenLaing}, with brightest spots within the extended lobes 
which are well back from the leading edges of
the lobes, and often in luminous jets. Three of these sources (3C\,28, 3C\,346
\& 3C\,348) have detected extended X-ray emission surrounding them 
\citep[]{HW99,GizLea,WB00}
so that for these cases, there is both direct and 
indirect evidence for a dense ICM. Additional
observations of AGN clusters with CHANDRA will be able to test directly 
the inferences made here (see Section
6.2).

6.  Correlations between redshift and $FR$ ratio and between redshift 
and projected physical size support an
ICM formation at $z \sim 0.4$ in environments of $B_{gg} \sim$ 
500-1000 Mpc$^{1.77}$.

7. Numerous X-ray selected clusters are known to exist in the 
redshift range  of our sample. These clusters
have radio galaxy populations which are exclusively FR1 in morphology 
and power level \citep[e.g.,][]{chap5}.
However, these clusters are substantially richer than the 
environments of the sources in our radio galaxy and
quasar samples at the same redshift, indicating that the formation 
epoch of a dense ICM depends, as expected
theoretically, on cluster richness.   In fact, the poorest of the 
EMSS cluster environments are comparable to
the richest galaxy environments in the radio galaxy and quasar samples.



Thus, the results of this study are consistent with the predictions 
of the ``evolutionary hypothesis'' for
radio-loud AGN first suggested by EYG. Specifically, in EYG and in
Paper 1, evidence was presented
that quasars found in clusters at $z\sim$0.5 fade quickly 
(``e-fading'' timescale of 0.9 Gyrs for H$_0$=50 km
s$^{-1}$ Mpc$^{-1}$) to become, first FR2 radio galaxies at 
$z\sim$0.25, and then FR1 radio galaxies at the
current epoch. While the physical mechanism for this fading is 
unknown, in Paper 1 we speculated that if
radio-loud AGN are powered by a rapidly spinning Black Hole
\citep[e.g.,][]{BZ77,BBR80,WC95}, the development of a deep 
gravitational potential and a dense ICM
around these AGN would prevent further  ``spin up'' by preventing 
the formation of new, supermassive Black
Hole binaries. These  binaries would not form because any 
galaxy-galaxy collisions in this cluster would be
both at much higher relative velocities than previously and also 
would be relatively gas free, so that bound
binary formation would be much less likely. In the absence of 
additional mechanisms to spin up the supermassive
Black Hole, each radio outburst would extract spin energy that could 
not be replaced and so each consecutive
outburst would be less powerful. Since the suggested timescale for radio 
outbursts ($\sim$10$^8$ years; Begelman et~al.\ 1984) is short
compared to the AGN fading timescale measured in Paper 1, the radio 
source power and structure would ``track'' the 
fading. So, while this scenario is consistent with a duty cycle of $\sim10\%$, 
our data do not address directly the question of AGN duty cycle. 
And, while the spin down of a supermassive Black Hole seems to us to be the most attractive 
scenario to account for the fading of cluster AGN, it is also possible to imagine that
an accretion powered AGN could have its fueling stifled by the development of a 
dense ICM \citep{StocPerr}.
 
\subsection{Prediction for Future Observations}

The important aspects of our conclusions which are amenable to test 
currently include the presence or absence
of a dense ICM based upon the FR2 radio source morphology and the 
possibility that FR1s in clusters were FR2s
in the past. In both cases these tests involve CHANDRA imaging spectroscopy.

The first prediction is in two parts. First, we predict that a dense 
X-ray emitting ICM will  be found around
the ``fat doubles'' at
$z\leq$0.4, some of which are shown in Figure 6. In the cases of 
3C\,348, 3C\,346 and 3C\,28 there is already
considerable evidence for extended X-ray emission around these three 
AGN \citep[e.g.,][] {GizLea}. Two other AGN
with small
$FR$ ratios at slightly higher $z$ (the quasar 3C\,215 with $FR$=0.65 at
$z$=0.411 and the radio galaxy 3C\,295 with
$FR$=0.78 at $z$=0.461) also have X-ray  evidence from ROSAT for a 
dense ICM as their radio morphologies and
$B_{gg}$ values (1000 and 1030 Mpc$^{1.77}$ respectively; see Paper 
1) predict. In fact,
\citet{Harris} have already detected  extended cluster X-ray emission 
around 3C\,295 with CHANDRA.  The other
``fat doubles'' should also be imaged with CHANDRA to make sure that 
the presence of an ICM is generic to
this class of sources.  While necessary, this first prediction 
is  insufficient to test completely our
use of radio morphology to locate a dense ICM. It is also important 
to verify that there is no ICM present
around quasars and FR2s of ``classical double'' morphology (i.e., 
high $FR$ ratios). In our survey this type
of source is found both in low-$B_{gg}$ regions at all redshifts and 
in high-$B_{gg}$ regions at $z>$0.4. Of
particular importance are the quasars with high-$B_{gg}$ at high-$z$, 
which are candidates for clusters whose
ICM has yet to form. Examples of this class include the quasars 
3C\,263 ($z$=0.646; $B_{gg}$=993
Mpc$^{1.77}$), 3C\,275.1 ($z$=0.557;
$B_{gg}$=1125 Mpc$^{1.77}$)  and PKS 0155-109 ($z$=0.616; 
$B_{gg}$=777 Mpc$^{1.77}$). 3C\,275.1 is
particularly important to observe with CHANDRA both because it has a 
large $\beta$ \citep[see map in][]{SBC},
and because there is a tentative ICM detection made with the 
ROSAT HRI. Our analysis in this paper
suggests that, despite the large $\beta$ and the large $B_{gg}$ value 
(Abell richness class 1), CHANDRA
observations will fail to confirm the tentative HRI detection of a 
dense cluster ICM around 3C\,275.1.
Since one definite CHANDRA detection already has been made 
(Worrall et~al.\ 2001)
of a dense ICM around a ``classical double'' FR2 at high-$z$
(3C\,220.1, $z$=0.620, $B_{gg}$=418 Mpc$^{1.77}$),
a few others would call into question
the methodology used herein to infer the presence or absence of a 
dense ICM from the radio source morphology.

Perhaps the most controversial hypothesis put forward in this paper 
and Paper 1 is the idea that FR2s in
clusters at high-$z$, fade and become FR1s in current epoch clusters. 
However, the recent  CHANDRA discovery
of ``holes'' in the ICM X-ray emission in some clusters  may offer a 
means of testing this hypothesis by
discovering ``fossil'' evidence of FR2s around FR1s.  In at least one 
case of an X-ray ``hole'' around the FR1
radio galaxy in Abell 4059 \citep{Heinz}, the inferred power required 
to evacuate  the ``hole'' of X-ray
emitting gas by $pdV$ work  is much larger than the power inferred to 
be present in the FR1
($\sim$5$\times$10$^{42}$ ergs s$^{-1}$) averaged over its 10$^8$ yr 
lifetime (Reynolds, Heinz \& Begelman 2001, which uses the
prescriptions in Bicknell, Dopita \& O'Dea 1997).
The X-ray cavity walls also do not correspond 
with the current boundaries of the radio
source lobes. Both the anomalously large size and power 
requirements of the X-ray cavity  in Abell 4059
suggest that the radio source was both larger and more powerful in 
the recent past; i.e., the previous
outburst was an FR2. Also, on the basis of the interpretation put 
forward in this paper, the ICM of Abell 4059
could have been much less dense in the recent past than now, also 
making the cavity easier to create.  On the
basis of the current work,  we predict that this one case is not 
unique, but that other, similar examples will
be found with CHANDRA. Indeed, the ``fat doubles'' in our sample are 
ideal targets to search for such
evidence since, from the ``evolutionary hypothesis'',  the most 
recent outburst of these sources is the first
after a dense ICM has formed around them.

\acknowledgments

M.H. acknowledges the support of a NASA Graduate Student Research 
Program Fellowship NGT - 51291 and publication
support from the National Radio Astronomy Observatories (NRAO).
H.K.C. Yee is thanked for providing the $B_{gg}$ values for some 
EMSS clusters prior to publication. J.P.
Leahy is thanked for allowing us to present his unpublished radio 
maps.  Everyone providing a ``private
communication'' listed in Table 1 is thanked for furnishing us with 
at least one unpublished radio map. W.J.M.
van Breugel is thanked for the unrestricted access to his supply of 
unpublished radio maps and data.  Michael
Rupen and the staff at NRAO are thanked for their AIPS support.  This 
research has made use of the NASA/IPAC
Extragalactic Database (NED) which is operated by the Jet Propulsion 
Laboratory, California Institute of
Technology, under contract with NASA.

\clearpage

%
%

%

\clearpage

\begin{figure}
\epsscale{1.0}
\plotone{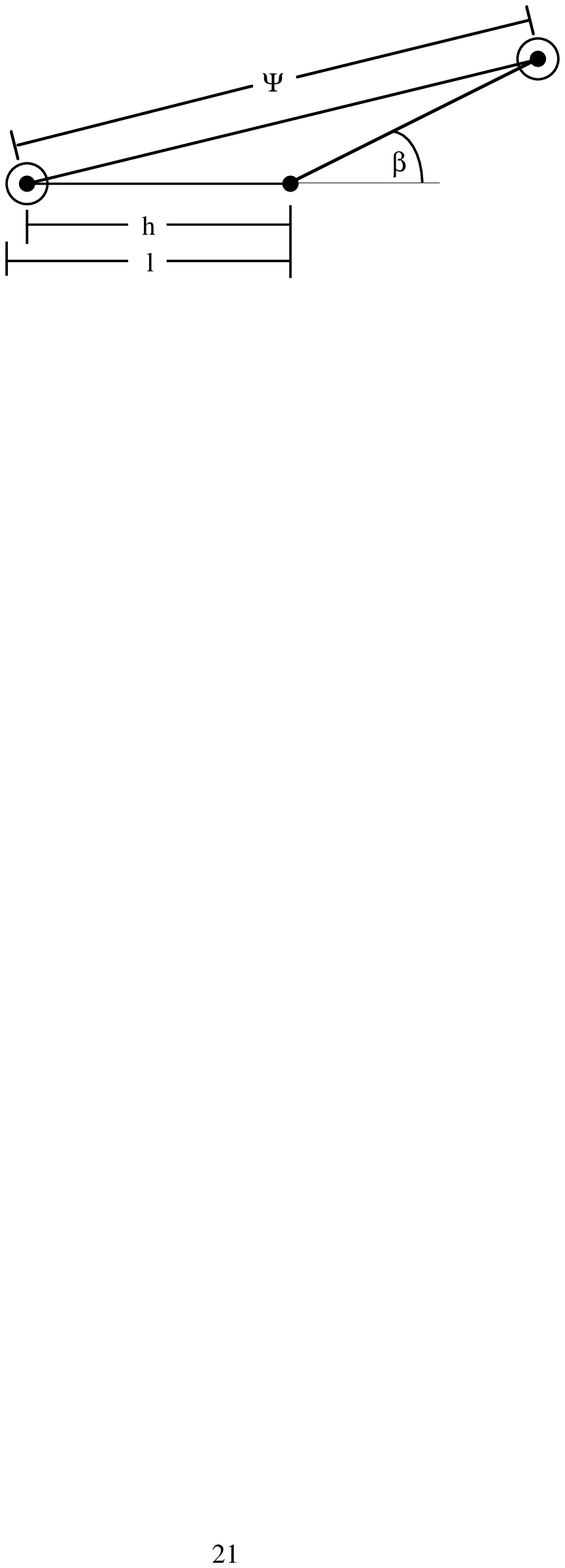}
\figcaption{The definitions of the peak-to-peak source size $\Psi$,
the bending angle $\beta$, and the $h$ and $l$ quantities used in the
calculation of the lobe length asymmetry $Q$, and the $FR$ ratio.
This figure was provided courtesy of Rector, Stocke \& Ellingson
(1995).}
\label{fig1}
\end{figure}

\begin{figure}
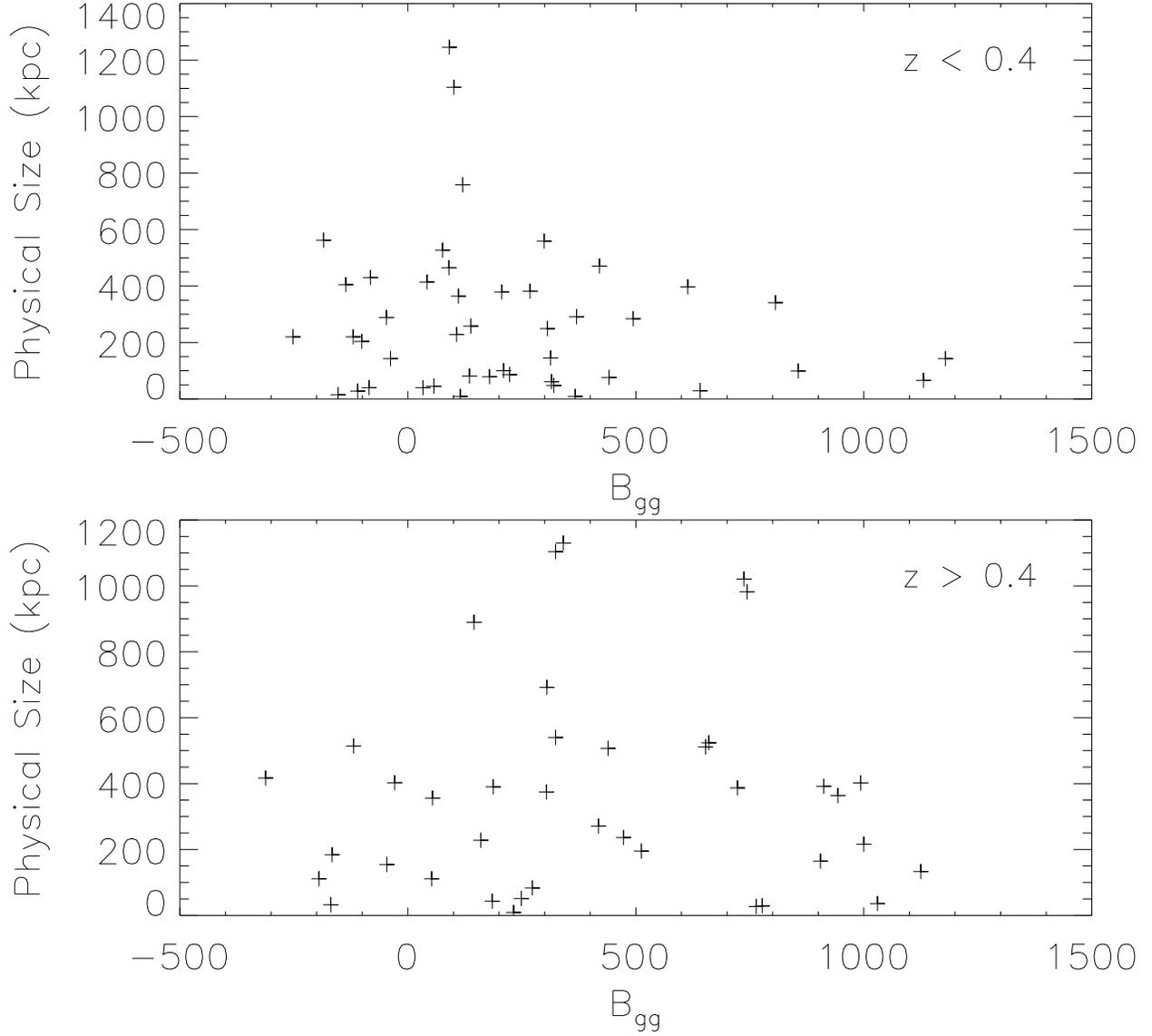

\epsscale{1.0}
\plotone{f2top.epsi}
\plotone{f2bot.epsi}
\figcaption{Projected physical size vs. environment ($B_{gg}$)
for the $z < 0.4$ (upper panel) and $z > 0.4$ (lower panel)
subsamples.  The correlation coefficient for the low z data
is $r = -0.157$; that for the high z data is $r = 0.037$.}
\label{fig2}
\end{figure}

\begin{figure}
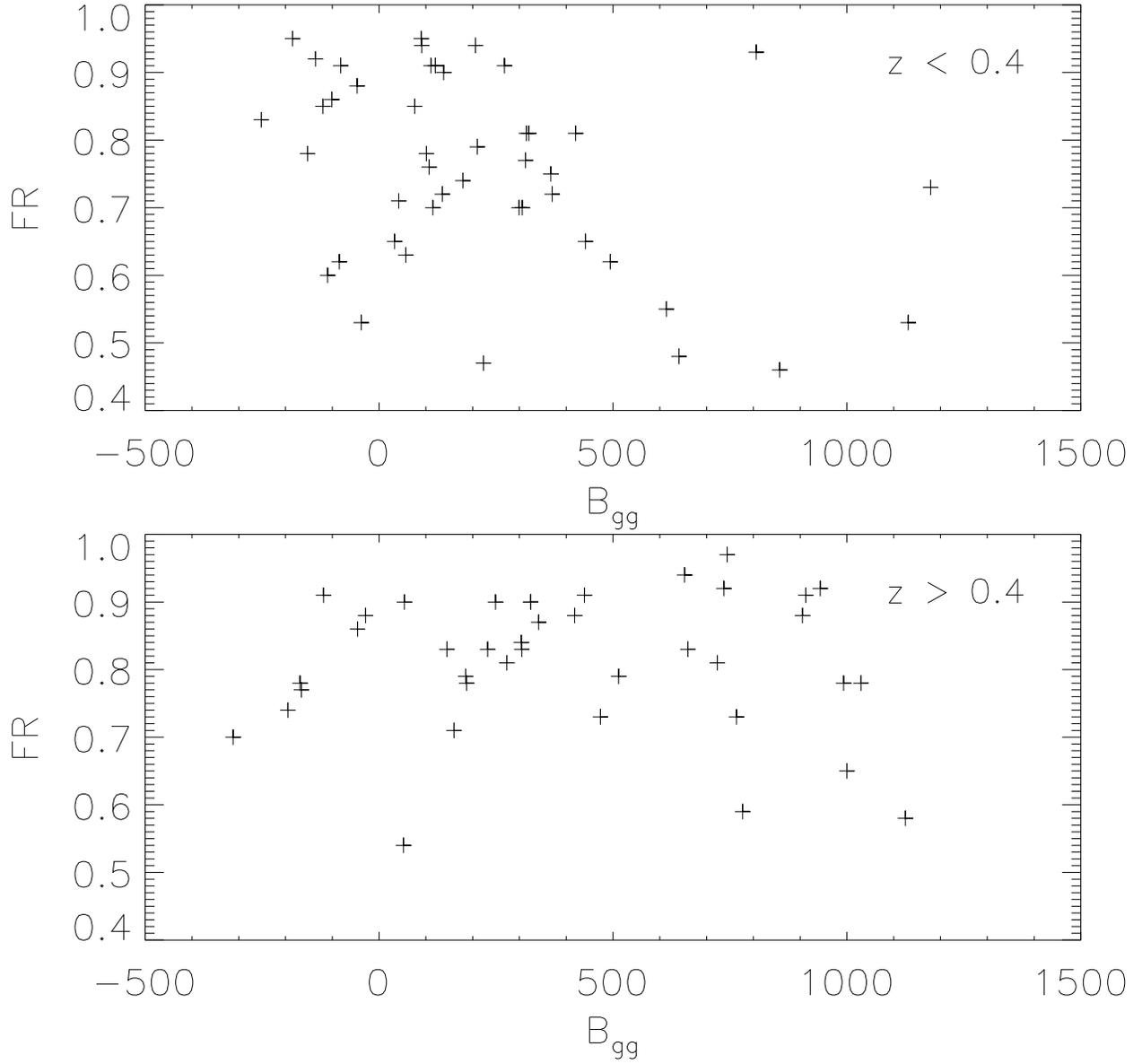

\epsscale{1.0}
\plotone{f3top.epsi}
\plotone{f3bot.epsi}
\figcaption{FR ratio ($FR$) vs. environment ($B_{gg}$)
for the $z < 0.4$ (upper panel) and $z > 0.4$ (lower panel)
subsamples.  The correlation coefficient for the low z data
is $r = -0.392$; that for the high z data is $r = -0.019$.}
\label{fig3}
\end{figure}

\begin{figure}
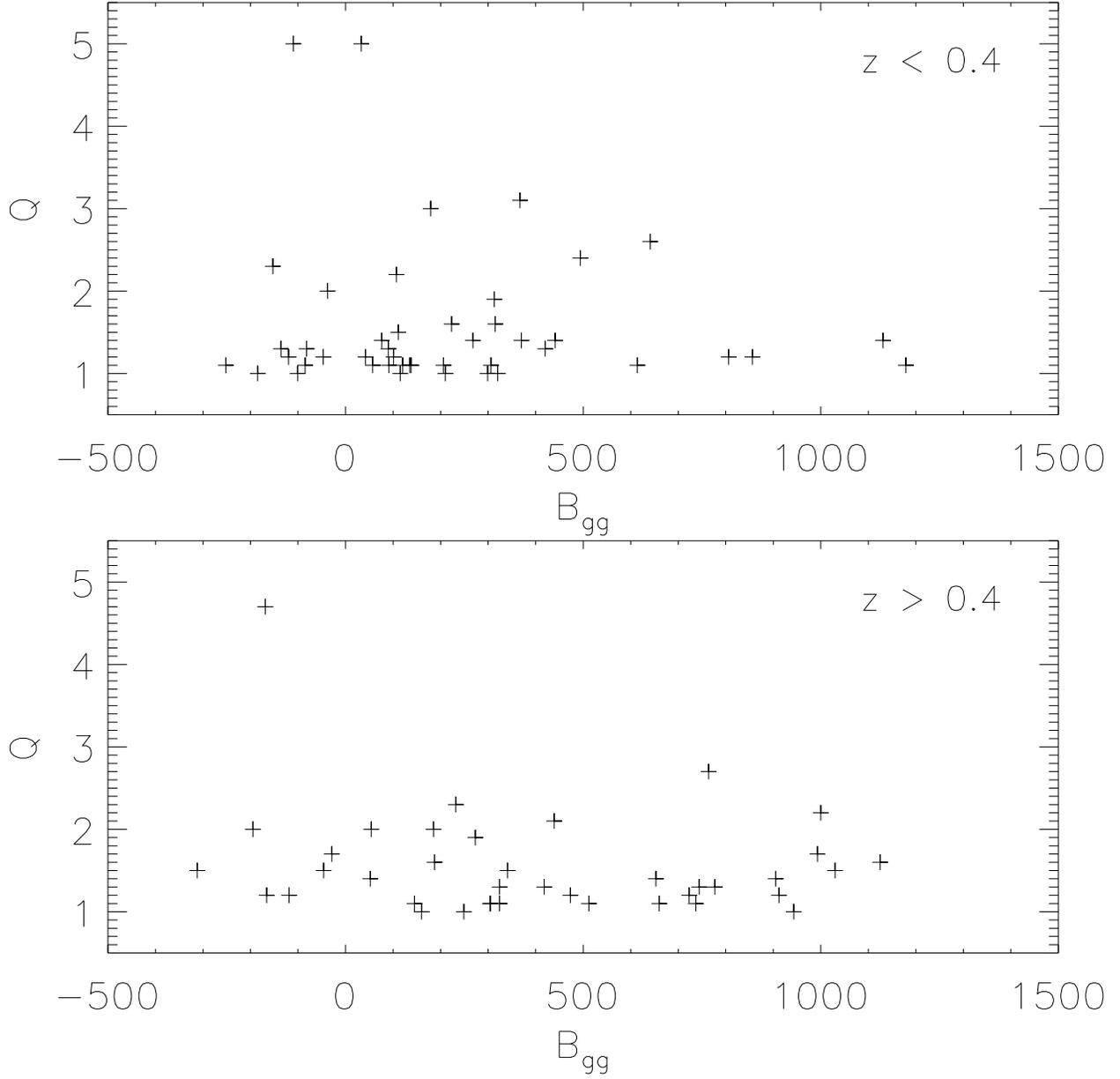

\epsscale{1.0}
\plotone{f4top.epsi}
\plotone{f4bot.epsi}
\figcaption{Lobe length asymmetry ($Q$) vs. environment
($B_{gg}$)
for the $z < 0.4$ (upper panel) and $z > 0.4$ (lower panel)
subsamples.  The correlation coefficient for the low z data
is $r = -0.091$; that for the high z data is $r = -0.193$.}
\label{fig4}
\end{figure}

\begin{figure}
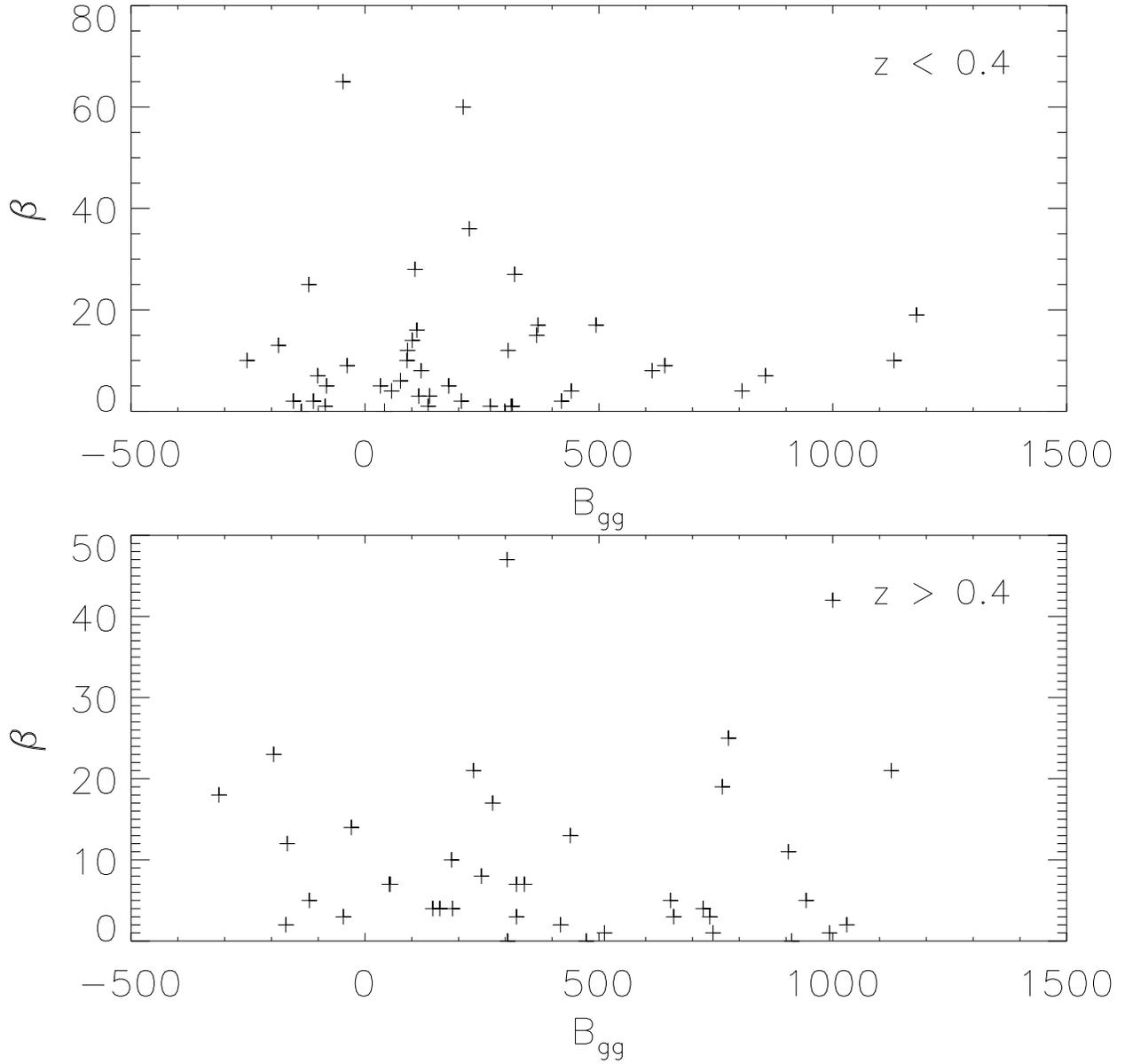

\epsscale{1.0}
\plotone{f5top.epsi}
\plotone{f5bot.epsi}
\figcaption{Projected bending angle ($\beta$; in degrees)
vs.  environment ($B_{gg}$)
for the $z < 0.4$ (upper panel) and $z > 0.4$ (lower panel)
subsamples.  The correlation coefficient for the low-z data
is $r = -0.008$; that for the high-z data is $r = 0.015$.}
\label{fig5}
\end{figure}

\begin{figure}
\epsscale{0.77}
\plottwo{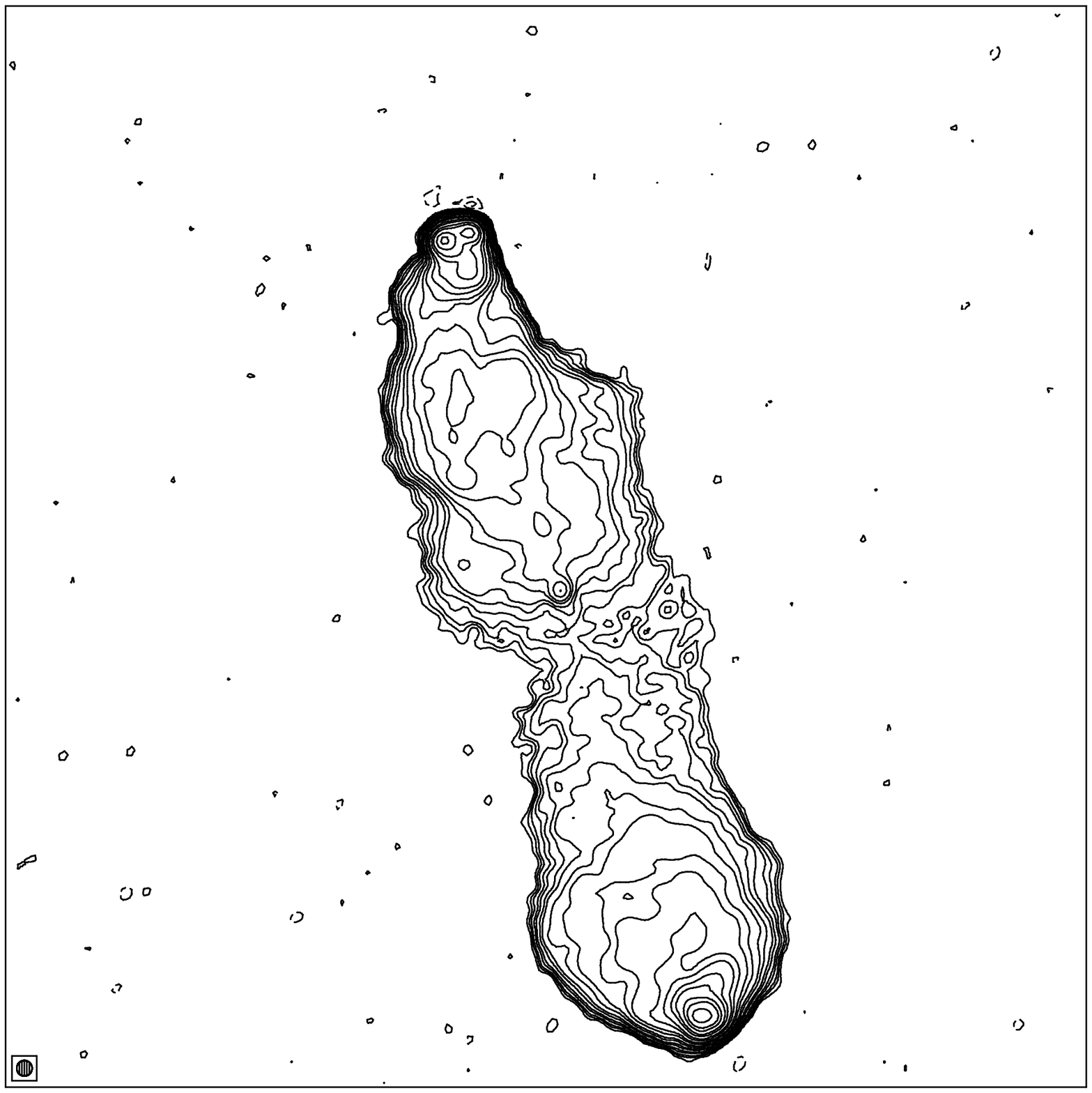}{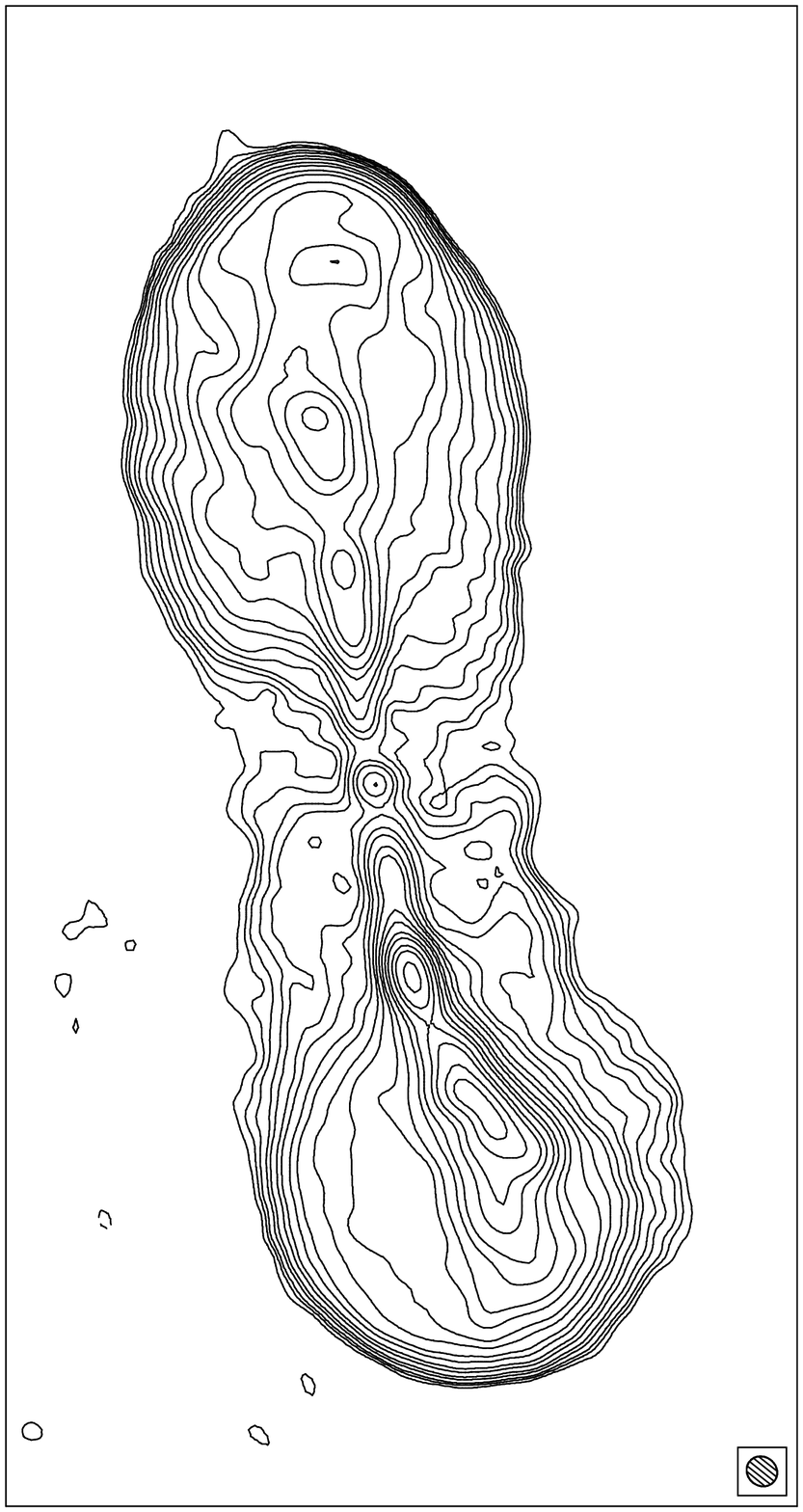}
\end{figure}
\begin{figure}
\epsscale{0.77}
\plottwo{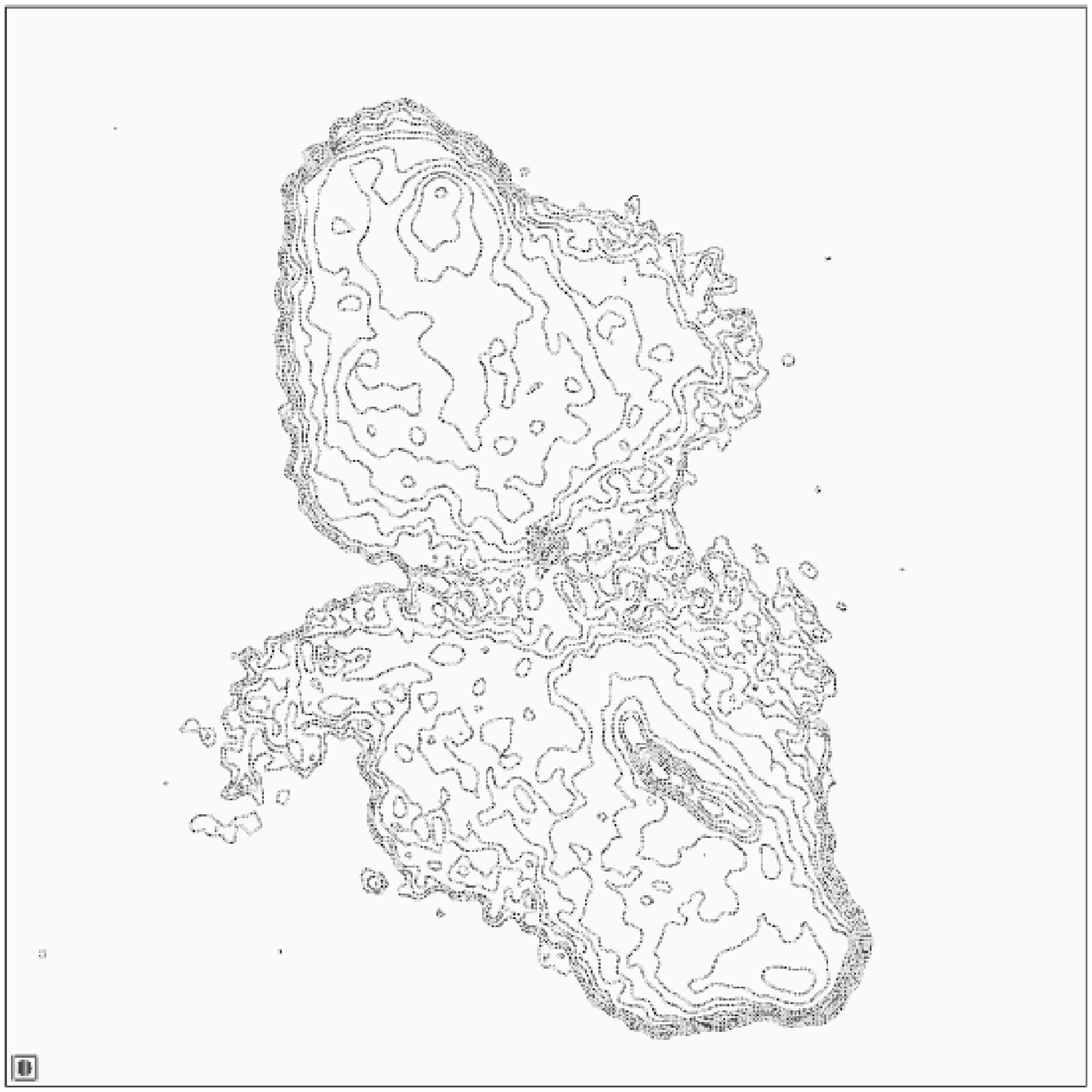}{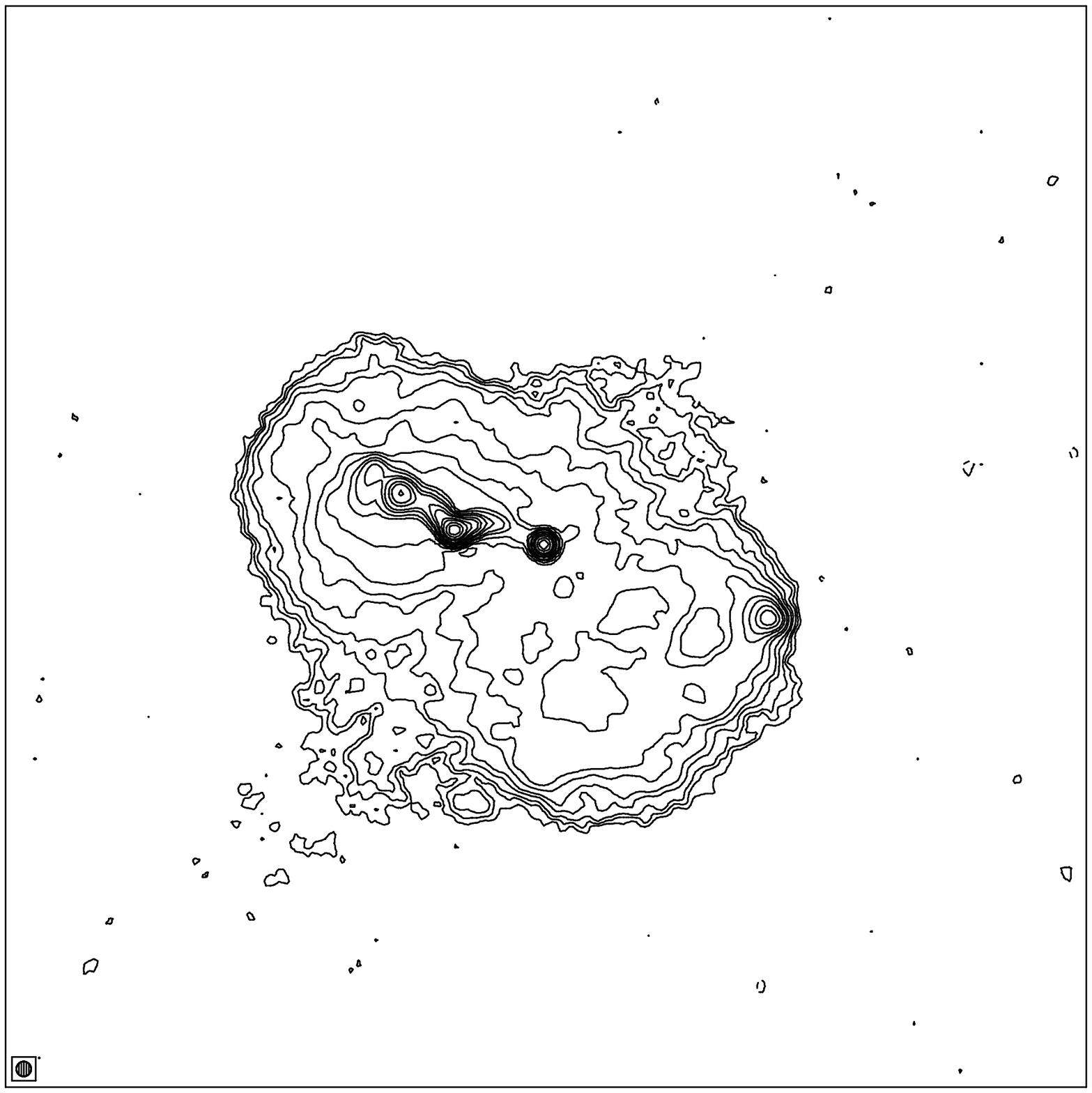}
\figcaption{Radio contour maps of sources with
$0.15 < z < 0.4$ in rich environments
($B_{gg} > 500$ Mpc$^{1.77}$).  Upper left panel:
3C\,173.1 (Leahy \& Perley 1991), $z=0.292$, $B_{gg}=806$.
Upper right panel: 3C\,348 (Harvanek \& Hardcastle 1998),
$z=0.154$, $B_{gg}=614$.  Lower left panel: 3C\,401
(J.P. Leahy unpublished), $z=0.201$, $B_{gg}=1131$.
Lower right panel: 3C\,346 (J.P. Leahy in prep.;
Akujor \& Garrington 1995), $z=0.161$, $B_{gg}=641$.
All maps are L band maps.  The beam for each map is
shown in one of the map corners.  The maps of 3C\,173.1,
3C\,401 and 3C\,346 were taken from the database of
Leahy, Bridle \& Strom (1997).  Note that all of these
sources except 3C\,173.1 show some radio structure
features that are characteristic of FR1 type radio
sources.}
\label{fig6}
\end{figure}

\begin{figure}
\epsscale{0.65}
\plottwo{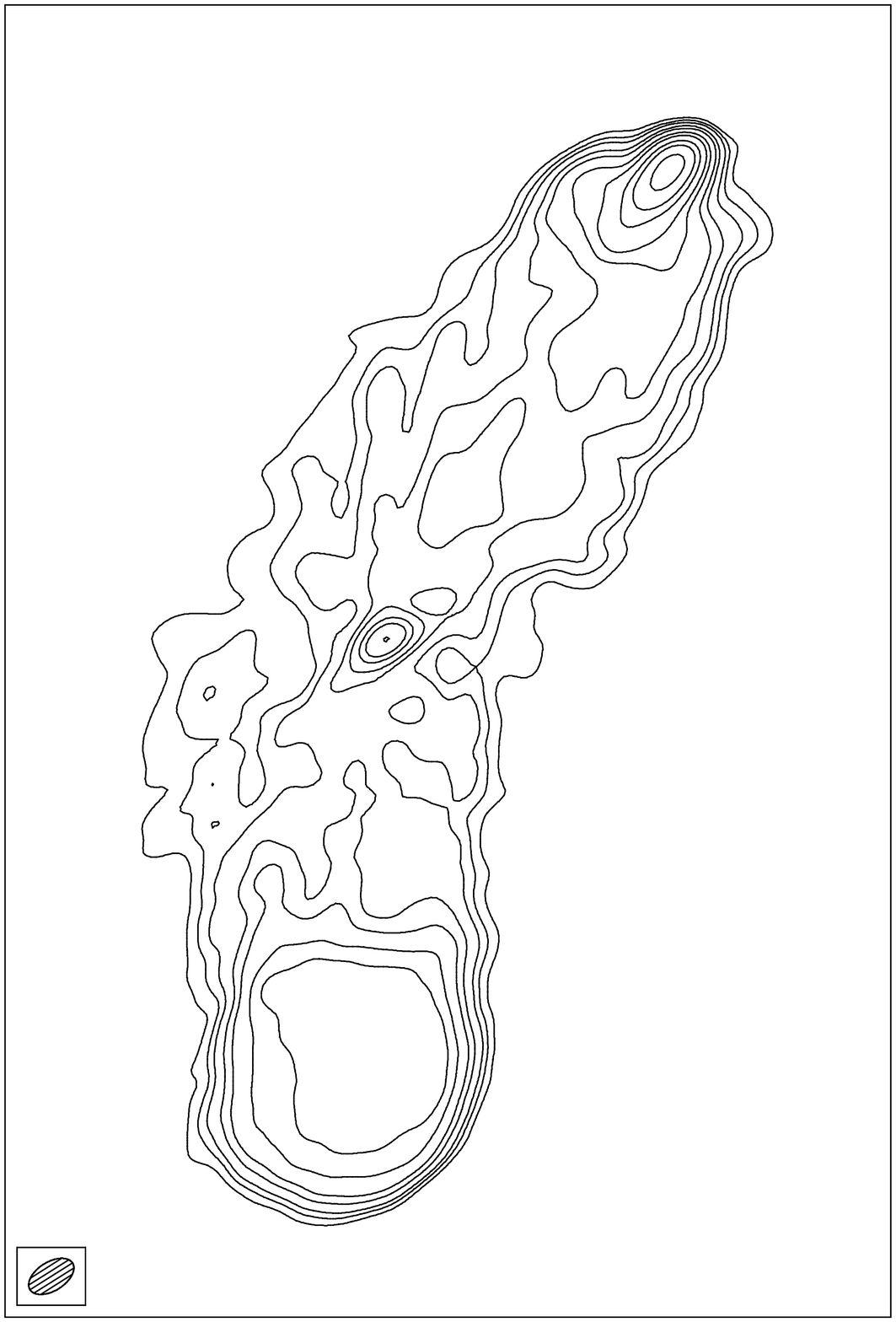}{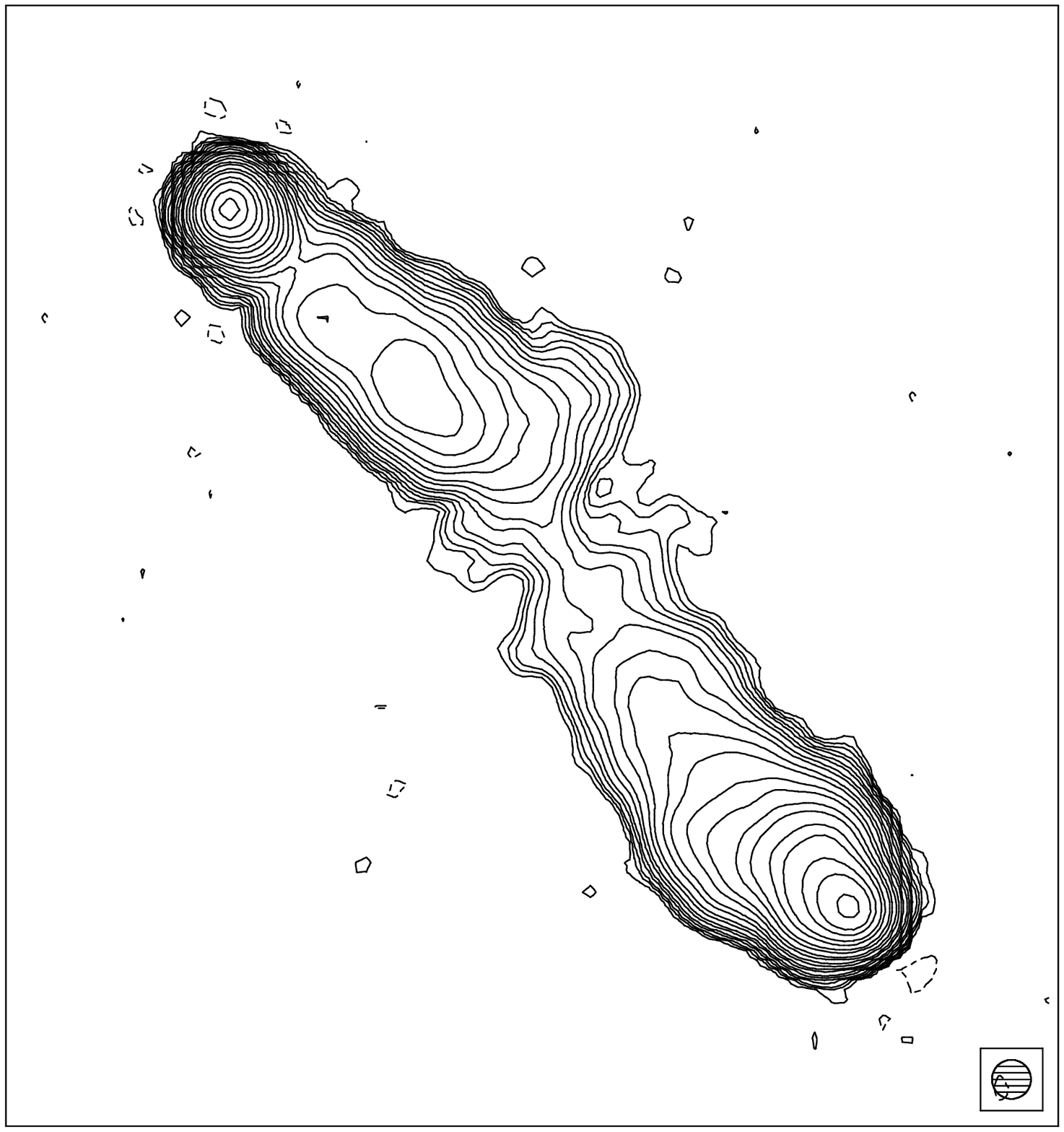}
\end{figure}
\begin{figure}
\epsscale{0.45}
\plottwo{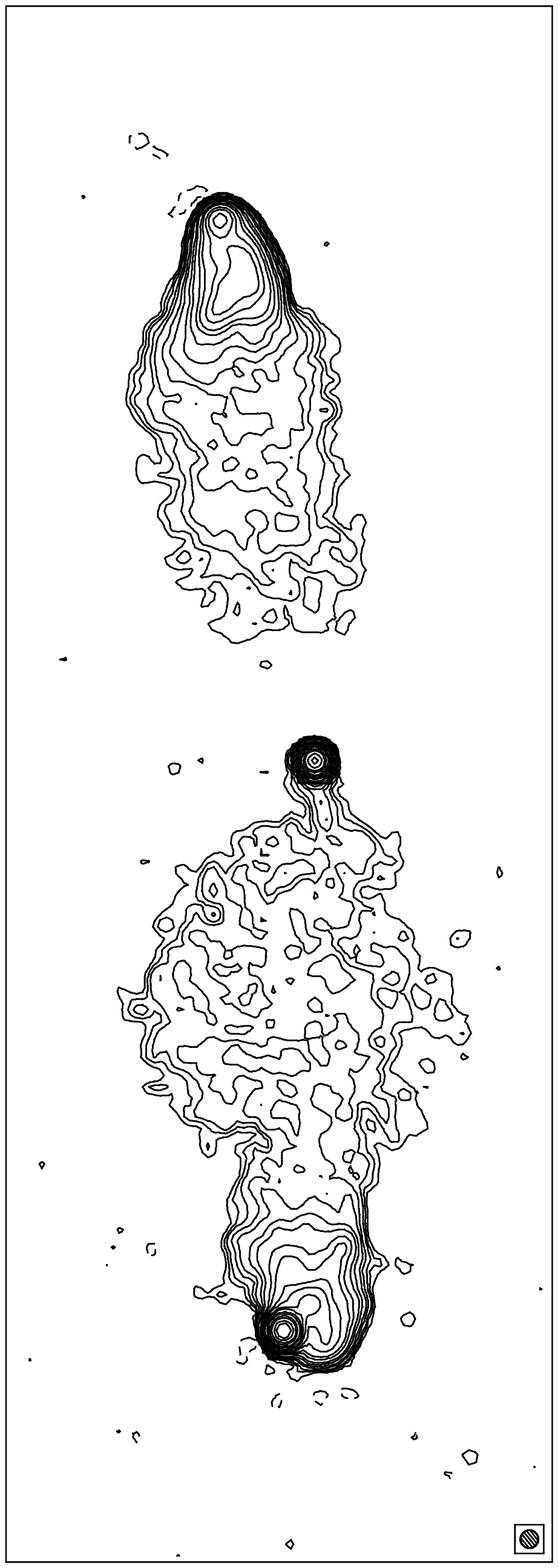}{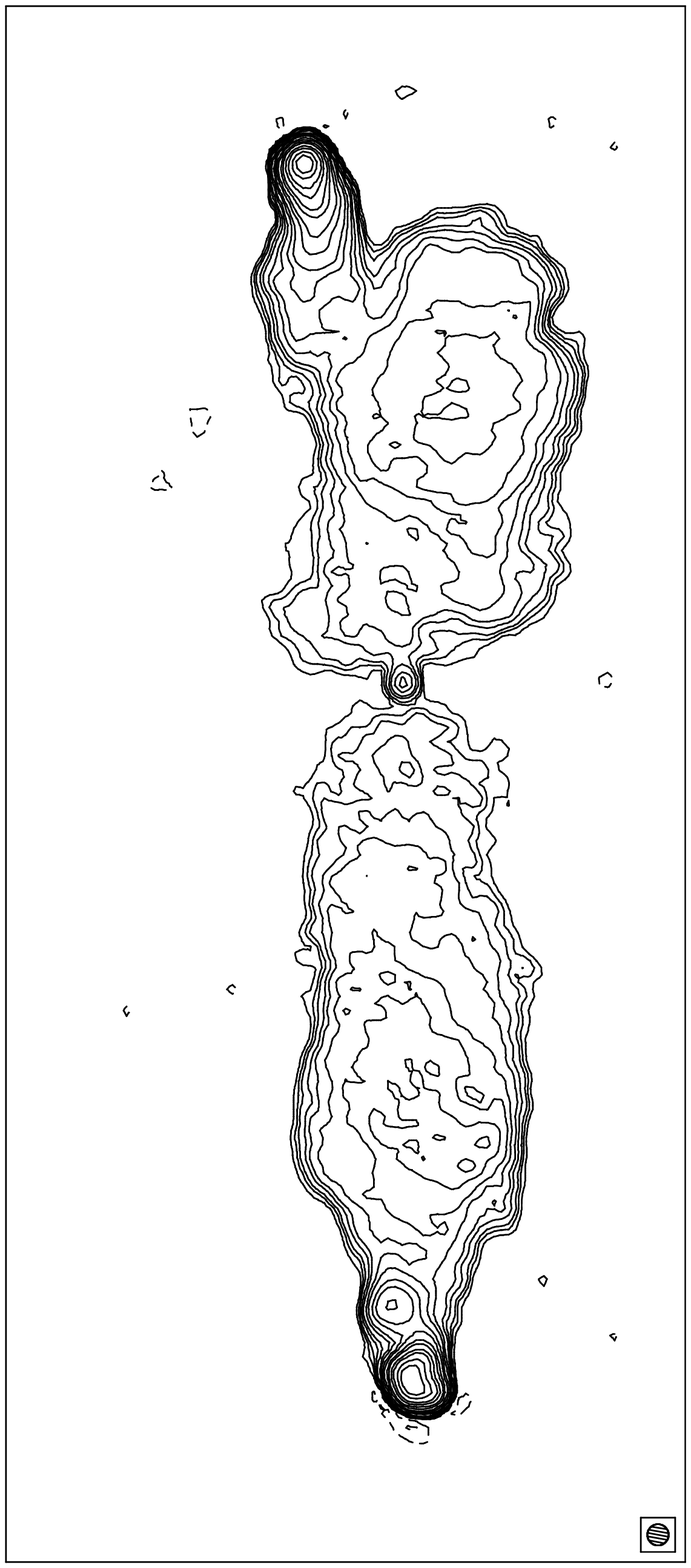}
\figcaption{Radio contour maps of sources with
$0.15 < z < 0.4$ in poor environments
($B_{gg} \sim 0$ Mpc$^{1.77}$).  Upper left panel:
3C\,18 (J.T. Stocke, unpublished), $z=0.188$,
$B_{gg}=-120$.  Upper right panel: 3C\,42 (Leahy \&
Perley 1991), $z=0.395$, $B_{gg}=-101$.  Lower left panel:
3C\,109 (Giovannini et~al.\ 1994), $z=0.3056$, $B_{gg}=-185$.
Lower right panel: 3C\,79 (Spangler, Myers \& Pogge 1984),
$z=0.2559$, $B_{gg}=90$.
All maps are L band maps.  The beam for each map is
shown in one of the map corners.  The maps of 3C\,42,
3C\,109 and 3C\,79 were taken from the database of
Leahy, Bridle \& Strom (1997).  Note that all of these
sources have a radio structure consistent with that of
an FR2 type radio source.}
\label{fig7}
\end{figure}

\begin{figure}
\epsscale{1.11}
\plottwo{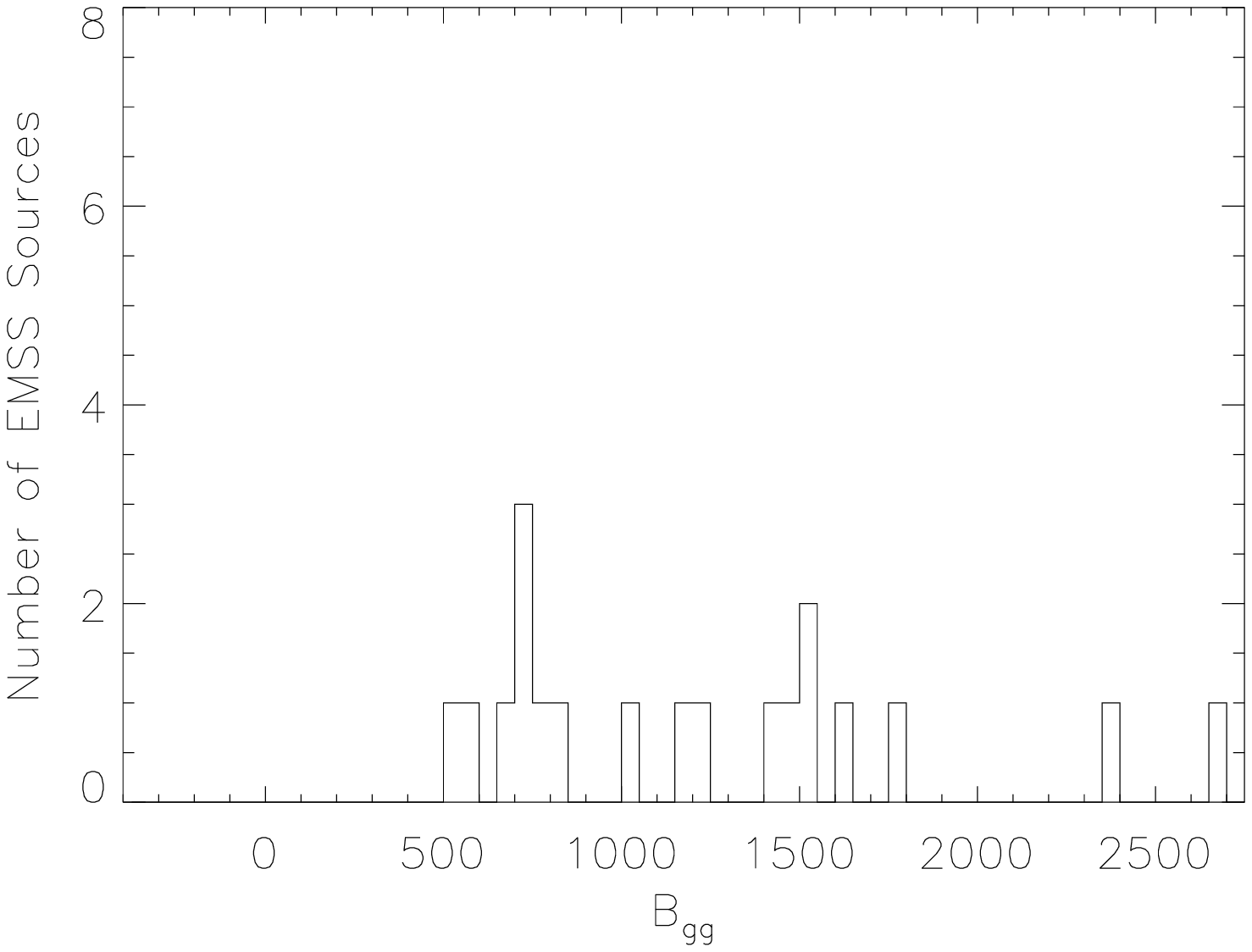}{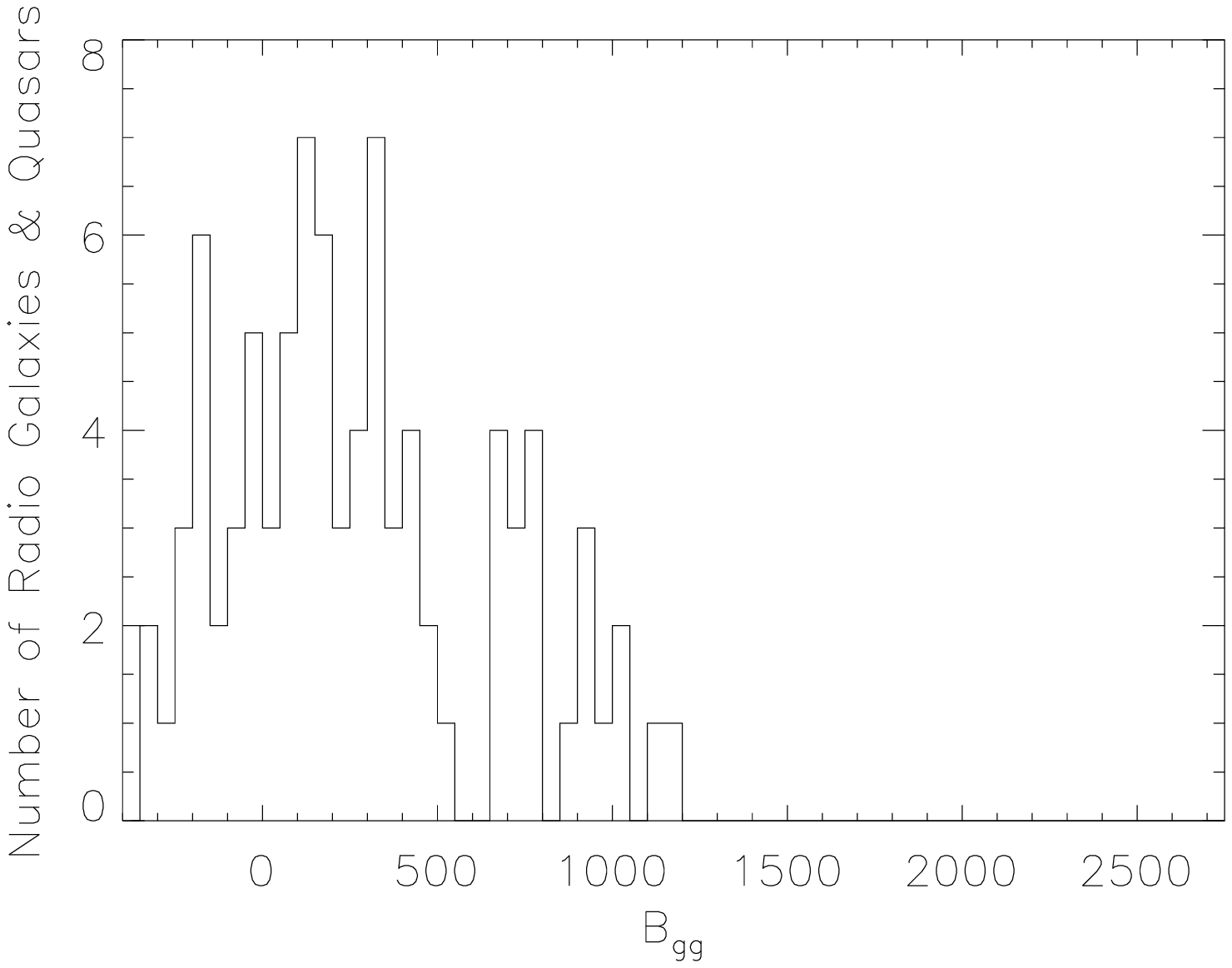}
\end{figure}
\begin{figure}
\epsscale{1.11}
\plottwo{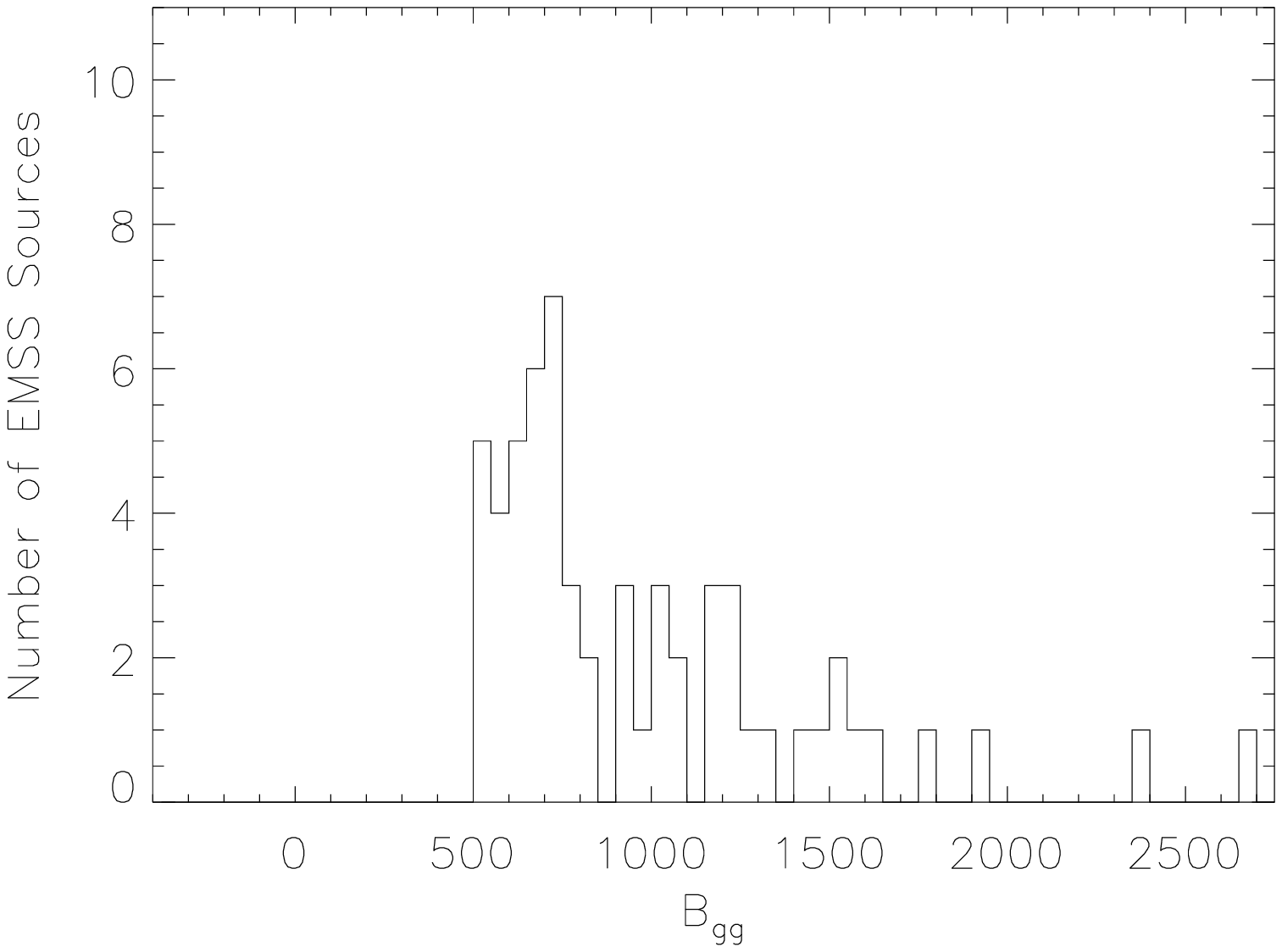}{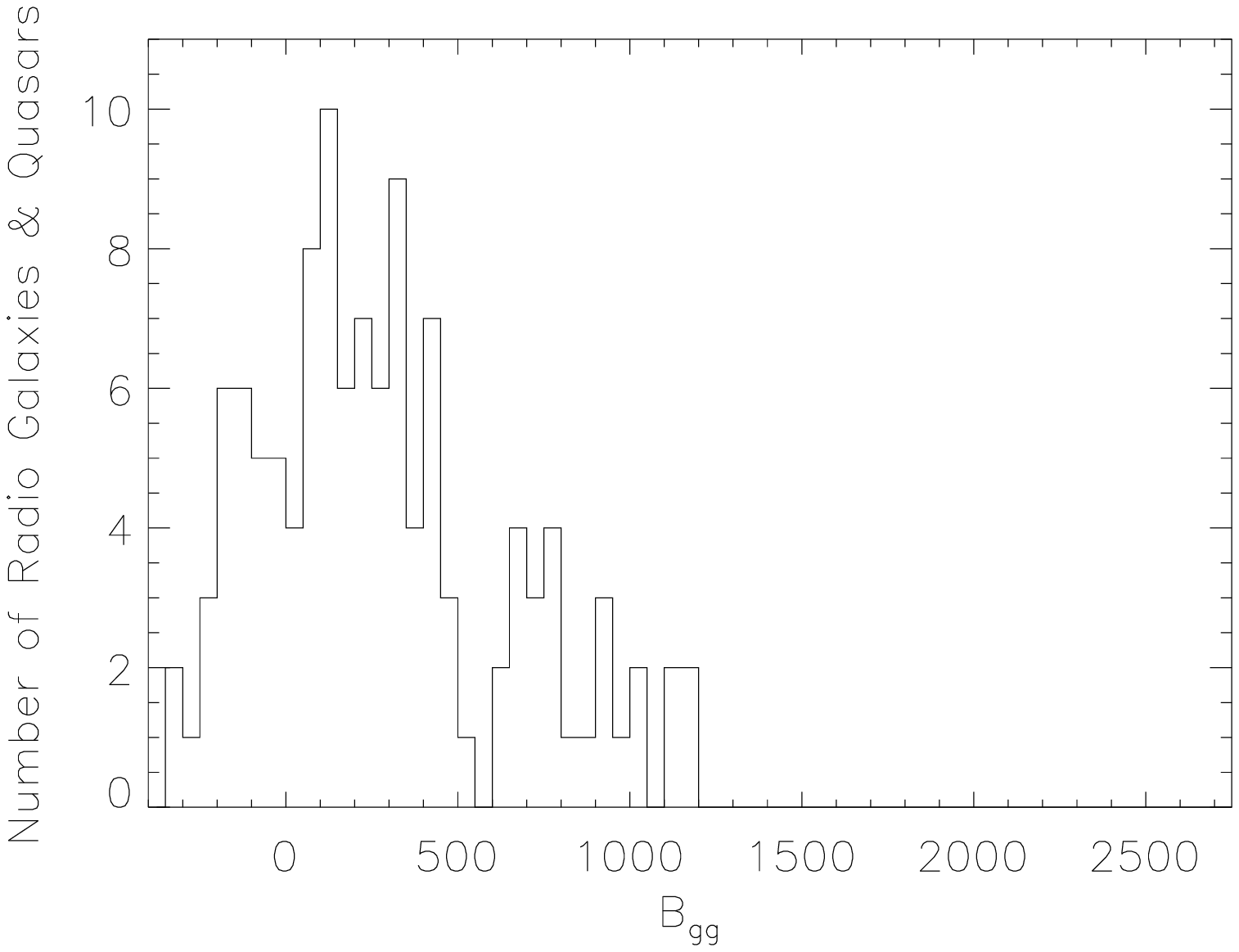}
\figcaption{Distributions of $B_{gg}$ values for an X-ray selected
and an AGN selected cluster sample. At top
are clusters at $z >$ 0.3; at bottom are clusters at
$z >$ 0.15. (a) At top left is the EMSS cluster sample of Stocke
et~al.\ (1999) (b) At top right is the $z > 0.3$ portions of the
radio galaxy and quasar samples studied herein;
(c) At bottom left is the $z >$ 0.15 EMSS cluster sample from
Gioia \& Luppino (1994); and (d) At bottom right is the full
radio galaxy + quasar sample studied herein.}
\label{fig8}
\end{figure}

\begin{table}
\caption{Radio Structure Parameters}
\label{radstruc}
\tiny
\begin{tabular}{ccccccccccccc}
\tableline \tableline Source & $z$ & LAS & Size & Ref., & $\Psi$ & 
Size & $\beta$ & Ref., & $Q$ & $FR$ & Ref., & Other Radio Refs. \\
(3CR\#) & & ($''$) & (kpc) & Band & ($''$) & 
(kpc) & ($^\circ$) & Band & & & Band & \\
(1) & (2) & (3) & (4) & (5) & (6) & 
(7) & (8) & (9) & (10) & (11) & (12) & (13) \\ \tableline
   16.0 & 0.406  &  80.3 &  577 & 41,X    &  70.7 &  507  &  13: & 
62,L     & 2.1: &  0.91: & 62,L &
62,83,46,12,60,41 \\
   17.0 & 0.2197 &  60.0 &  286 & 80,C    &  18.0 &   86  &  36 & 
107,C    & 1.6* &  0.47$\dag$ & 107,L &
107,80,116 \\
   18.0 & 0.188  &  65.8 &  279 & 80,C    &  51.9 &  220  &  25 & 
116,L     & 1.2 &  0.85 & 116,L & 107,80,116
\\
   19.0 & 0.482  &   7.5 &   59 & 73,X    &   6.5 &   51  &   8 & 60,L 
& 1.0 &  0.90 & 60,L & 46,73,60,116
\\
   28.0 & 0.1952 &  50.2 &  219 & 33,C    &  32.8 &  143  &  19: & 
58,L     & 1.1: &  0.73: & 58,L &
107,96,72,29,33,58,116,60 \\
   42.0 & 0.395  &  33.2 &  235 & 62,L    &  28.9 &  204  &   7 & 30,C 
& 1.0 &  0.86 & 62,L &
72,62,46,60,30 \\
   46.0 & 0.4373 & 174.4 & 1308 & 37,L    & 150.6 & 1130  &   7 & 60,L 
& 1.5 &  0.87 & 60,L &
71,46,37,60,82,119 \\
   47.0 & 0.425  &  79.3 &  585 & 59,L    &  69.7 &  514  &   5 & 59,L 
& 1.2 &  0.91 & 59,L &
71,89,14,21,31,90,16,60,59 \\
   48.0 & 0.367  &   $\ge$0.9 &    $\ge$6 & 121,L    &  ---  &  ---  & 
--- & 121,L     & --- &  ---  & &
3,118,83,117,105,90,94,101,60,121,1 \\
   49.0 & 0.621  &   $\ge$1.2 &   $\ge$11 & 99,C    &   1.0 &    9  & 
21 & 99,C     & 2.3 &  $\le$0.83 & 99,C
& 3,83,117,105,27,28,94,99,82 \\
   61.1 & 0.186  & 200.2 &  841 & 62,L    & 185.5 &  780  &   8 & 62,L 
& 1.2 &  0.92 & 62,L &
107,71,14,40,56,62,60,82,5 \\
   63.0 & 0.175  &  45.2 &  181 & 41,X    &  18.9 &   76  &   4 & 
116,C     & 1.4* &  $\le$0.65$\dag$ & 41,L &
10,116,41,95,113 \\
   67.0 & 0.3102 &   3.2 &   19 & 52,X    &   2.5 &   15  &   2 & 99,C 
& 2.3 &  0.78 & 99,L &
3,118,83,117,105,27,94,48,51,70,60,99,1,53,52,49 \\
   79.0 & 0.2559 &  91.1 &  485 & 102,L    &  87.3 &  465  &  10 & 
38,X     & 1.3 &  0.95 & 102,L &
107,96,71,102,7,38,60,39 \\
   93.0 & 0.358  &  $\ge$39.4 &  $\ge$262 & 12,L    &  33.0 &  220  & 
10 & 12,X     & 1.1 &  $\le$0.83 & 12,L
& 90,12,50 \\
   93.1 & 0.244  &   $\le$1.5 &    $\le$8 & 105,L    &  ---  &  ---  & 
--- & 105,L     & --- &  ---  & &
3,105,83,1 \\
   99.0 & 0.426  &   $\ge$5.8 &   $\ge$43 & 76,L    &   4.4 &   32  & 
2 & 75,X     & 4.7 &  $\le$0.78 & 76,L
& 76,73,82,75 \\
  109.0 & 0.3056 & 101.7 &  611 & 35,L    &  93.5 &  562  &  13 & 35,L 
& 1.0 &  0.95 & 35,L &
96,71,14,56,10,7,60,35 \\
  142.1 & 0.4061 &  $\ge$54.5 &  $\ge$391 & 12,L    &  49.6 &  356  & 
7 & 12,X     & 2.0 &  $\le$0.90 & 12,L
& 12,9,116 \\
  169.1 & 0.633  &  57.9 &  526 & 41,X    &  45.9 &  417  &  18 & 41,X 
& 1.5* &  0.70 & 41,L & 72,115,41,82
\\
  171.0 & 0.2384 &  35.1 &  178 & 107,L   &   9.0 &   45  &   4 & 42,U 
& 1.1 &  0.63$\dag$ & 42,L &
107,89,56,42,10,7,38,116,60,82,39,34 \\
  173.1 & 0.292  &  63.1 &  368 & 62,L    &  58.5 &  341  &   4 & 38,X 
& 1.2 &  0.93 & 62,L &
107,71,62,46,38,60,39,34 \\
  196.1 & 0.198  &  10.0 &   44 & 82,L    &  ---  &  ---  & --- & 82,L 
& --- &  ---  &  & 10,82 \\
  200.0 & 0.458  &  27.3 &  210 & 12,L    &  14.5 &  111  &   7 & 60,C 
& 1.4* &  0.54 & 12,L &
21,32,46,37,12,60 \\
  213.1 & 0.194  &  45.7 &  198 & 37,L    &   5.9 &   26  &  10 & 1,X 
& 2.2 &
$\le$0.19$\dag$ & 1,L & 107,3,105,37,1 \\
  215.0 & 0.411  &  63.6 &  460 & 32,L    &  29.9 &  216  &  42 & 16,C 
& 2.2* &  0.65$\dag$ & 45,L &
72,89,14,32,43,110,90,16,44,60,45
\\
  219.0 & 0.1744 & 205.1 &  820 & 119,L    & 147.9 &  591  &  10 & 
22,L     & 1.0 &  $\le$0.81$\dag$ & 22,L &
71,14,18,86,23,22,60,82,119 \\
  220.1 & 0.620  &  45.5 &  409 & 41,L    &  30.1 &  271  &   2 & 21,C 
& 1.3 &  $\le$0.88 & 21,C &
71,21,46,41 \\
  225.0B& 0.582  &   6.3 &   55 & 34,C    &   5.0 &   43  &  10: & 
34,C     & 2.0: &  0.79: & 34,C &
71,14,46,9,116,34 \\
  228.0 & 0.5524 &  51.1 &  435 & 47,L    &  46.1 &  392  &   0 & 47,L 
& 1.2 &  0.91 & 47,L &
72,21,46,116,47 \\
  234.0 & 0.1848 & 123.2 &  515 & 63,L    & 110.7 &  463  &  10 & 38,X 
& 1.4 &  0.89 & 63,L &
96,71,64,21,38,60,63,82,39,116,6,4 \\
  244.1 & 0.428  &  61.8 &  458 & 64,L  &  52.2 &  387  &   4 & 30,C 
& 1.2 &  0.81 & 64,L &
71,64,56,46,58,60,63,82,34,30,6 \\
  249.1 & 0.311  &  50.0 &  304 & 61,L    &  23.5 &  143  &   9 & 79,C 
& 2.0* &  0.53$\dag$ & 61,L &
72,89,21,79,61,69,16,60 \\
  263.0 & 0.646  &  59.8 &  548 & 45,L    &  43.9 &  402  &   1 & 16,C 
& 1.7* &  0.78 & 45,L &
71,89,61,16,12,93,45 \\
  268.2 & 0.362  & $\ge$113.2 &  $\ge$758 & 41,L    &  96.4 &  646  & 
6 & 82,L     & 1.0 &  $\le$0.85 & 41,L
& 96,71,109,41,82 \\
  268.3 & 0.371  &   1.9 &   13 & 27,L    &   1.4 &    9  &  15 & 70,C 
& 3.1 &  $\le$0.75 & 70,C &
3,118,83,105,27,94,70,60,1,82 \\
  273.0 & 0.158  &  25.2 &   93 & 24,L    &  ---  &  ---  & --- & 24,L 
& --- &  ---  & & 87,24,19,55 \\
  274.1 & 0.422  & 170.5 & 1253 & 63,L    & 150.2 & 1104  &   3: & 
64,L   & 1.1: &  0.90: & 64,L &
71,64,109,46,60,63,6 \\
  275.0 & 0.480  &   6.8 &   54 & 74,C    &   5.4 &   42  &  11: & 
74,C     & 2.4: &  0.78: & 74,C &
10,74,73,9,116,75 \\
  275.1 & 0.557  &  36.0 &  307 & 45,L    &  15.6 &  133  &  21 & 
108,C     & 1.6 &  0.58 & 68,L &
32,108,65,46,68,66,12,57,45 \\
  277.0 & 0.414  & $\ge$158.8 & $\ge$1154 & 41,L    & 131.9 &  958  & 
5 & 41,L     & 1.7 &  $\le$0.83 & 41,L
& 71,109,116,41 \\
  277.1 & 0.320  &   2.4 &   15 & 99,L    &   1.6 &   10  &   0 & 99,L 
& 2.5 &  0.63 & 99,L &
3,118,83,105,93,116,99,1 \\
  284.0 & 0.2394 & 190.5 &  966 & 63,L    & 174.4 &  885  &   2 & 38,X 
& 1.5 &  0.91 & 64,L &
107,96,71,64,37,38,60,63,39,6 \\
  287.1 & 0.2159 & $\ge$139.4 &  $\ge$656 & 41,L    & 112.1 &  527  & 
6 & 107,C    & 1.4 &  $\le$0.85 & 41,L
& 107,7,25,41 \\
  288.0 & 0.246  &  $\ge$36.0 &  $\ge$186 & 15,L    &  14.9 &   77  & 
17 & 15,C     & 1.5* &  $\le$0.46$\dag$
& 15,L & 89,15,98,68,66,60 \\
  295.0 & 0.4614 &   6.2 &   48 & 2,L    &   4.7 &   36  &   2 & 111,U 
& 1.5 &  0.78 & 88,X &
89,56,2,10,111,116,60,88 \\
  299.0 & 0.367  &  $\ge$12.9 &   $\ge$87 & 68,C    &  11.7 &   79  & 
5 & 60,L     & 3.0* &  0.74 & 68,L &
3,106,67,118,83,115,117,105,94,68,66,116,60,1 \\
  300.0 & 0.270  & 107.4 &  593 & 63,L    &  99.0 &  547  &   5 & 38,X 
& 2.4 &  0.88 & 64,L &
107,96,71,64,38,116,60,63,39,6 \\ \tableline
\end{tabular}
\end{table}
\begin{table}
\tiny
\begin{tabular}{ccccccccccccc}
\multicolumn{13}{l}{Radio Structure Parameters (continued)}\\
\tableline \tableline
Source & $z$ & LAS & Size & Ref., & $\Psi$ & 
Size & $\beta$ & Ref., & $Q$ & $FR$ & Ref.,
& Other Radio Re fs. \\
(3CR\#) & & ($''$) & (kpc) & Band & ($''$) & 
(kpc) & ($^\circ$) & Band & & & Band & \\
(1) & (2) & (3) & (4) & (5) & (6) & 
(7) & (8) & (9) & (10) & (11) & (12) & (13) \\ \tableline
  303.1 & 0.267  &   2.7 &   15 & 83,C    &   1.7 &    9  &   3: & 1,X 
& 1.0: &  0.70: & 1,X &
3,83,117,105,27,94,99,1 \\
  306.1 & 0.441  & 108.0 &  814 & 41,L    &  91.8 &  692  &   0 & 
116,C     & 1.1 &  0.83 & 41,L & 115,116,41 \\
  313.0 & 0.461  & $\ge$135.1 & $\ge$1044 & 10,C    & 127.1 &  982  & 
1 & 12,X     & 1.3 &  $\le$0.97 & 12,X
& 71,10,12,9 \\
  319.0 & 0.192  & 112.1 &  483 & 64,L  &  65.9 &  284  &  17: & 64,L 
& 2.4: &  0.62: & 64,L &
71,64,46,38,60,63,39,6 \\
  320.0 & 0.342  &  37.2 &  240 & 41,X    &  15.3 &   99  &   7 & 41,X 
& 1.2* &  0.46? & 37,L & 72,37,98,41
\\
  323.1 & 0.264  &  $\ge$76.2 &  $\ge$414 & 36,L    &  70.2 &  382  & 
1 & 79,C     & 1.4 &  $\le$0.91 & 12,L
& 71,89,14,79,110,36,12 \\
  327.1 & 0.4628 &  20.8 &  161 & 12,L    &  14.3 &  111  &  23 & 12,X 
& 2.0 &  0.74$\dag$ & 12,L &
10,80,12,116 \\
  330.0 & 0.550  &  73.8 &  626 & 61,L    &  61.7 &  524  &   3 & 30,C 
& 1.1* &  0.83 & 61,L &
71,56,61,77,46,58,30 \\
  332.0 & 0.1515 & 101.0 &  361 & 107,L   &  69.8 &  249  &  12 & 
107,C    & 1.1 &  0.70 & 107,L &
107,96,71,14,7,98,116 \\
  334.0 & 0.555  &  61.8 &  527 & 61,L    &  45.7 &  390  &   4 & 16,C 
& 1.6 &  0.78$\dag$ & 61,L &
71,32,61,43,46,90,16,12,45 \\
  337.0 & 0.635  &  $\ge$47.7 &  $\ge$434 & 85,L    &  42.6 &  388  & 
2 & 85,C     & 1.8 &  $\le$0.90 & 85,L
& 71,14,46,85,84,11 \\
  341.0 & 0.448  &  79.8 &  607 & 62,L    &  71.1 &  540  &   7 & 17,C 
& 1.3* &  0.90 & 62,L &
71,62,46,17,37,60 \\
  345.0 & 0.594  &  29.7 &  262 & 100,C    &  ---  &  ---  & --- & 
100,C     & --- &  ---  & &
83,20,54,114,13,19,100,45,120,122,91,92,55 \\
  346.0 & 0.161  &  17.5 &   66 & 117,L    &   7.8 &   29  &   9 & 1,L 
& 2.6* &  0.48 & 117,L &
107,89,3,106,117,10,116,60,1 \\
  348.0 & 0.154  & $\ge$201.7 &  $\ge$730 & 41,L    & 109.8 &  397  & 
8 & 41,L     & 1.1 &  $\le$0.55 & 41,L
& 26,116,41 \\
  349.0 & 0.205  &  89.4 &  404 & 38,X    &  83.9 &  379  &   2 & 38,X 
& 1.1 &  0.94 & 62,L &
107,71,56,62,46,38,60,39 \\
  351.0 & 0.371  &  76.2 &  518 & 62,L    &  63.2 &  430  &   5 & 62,L 
& 1.3 &  0.91$\dag$ & 62,L &
96,72,14,56,79,61,62,90,16,93,60 \\
  357.0 & 0.1664 & $\ge$117.6 &  $\ge$453 & 41,L    &  75.5 &  291  & 
17 & 41,C     & 1.4 &  $\le$0.72$\dag$ &
41,L & 107,96,71,116,41 \\
  379.1 & 0.256  &  $\ge$88.1 &  $\ge$469 & 104,C    &  76.1 &  405  & 
0 & 104,L     & 1.3 &  $\le$0.92 &
104,L & 107,71,40,56,104,81 \\
  381.0 & 0.1605 &  83.5 &  312 & 62,L    &  68.9 &  258  &   3 & 62,L 
& 1.1 &  0.90 & 62,L &
107,96,71,14,56,62,7,38,60,39 \\
  401.0 & 0.201  &  26.4 &  118 & 41,L    &  14.7 &   66  &  10 & 60,L 
& 1.4* &  0.53 & 41,L &
72,14,21,56,46,38,60,41,39 \\
  411.0 & 0.467  &  31.9 &  248 & 103,L    &  27.2 &  211  &   2 & 
82,C     & 1.0 &  0.87 & 82,L & 89,103,82 \\
  427.1 & 0.572  &  32.8 &  284 & 61,L    &  21.2 &  184  &  12 & 57,C 
& 1.2 &  0.77 & 82,L &
71,89,14,56,61,58,57,82 \\
  434.0 & 0.322  &  21.3 &  132 & 41,X    &  13.1 &   81  &   1 & 41,L 
& 1.1 &  0.72 & 41,L & 89,41 \\
  435.0A& 0.471  &  15.5 &  121 & 78,L  &  11.5 &   90  &  11 & 82,C 
& 1.4 &  0.74 & 78,L & 72,78,116,82,97
\\
  436.0 & 0.2145 & 127.1 &  595 & 113,L    & 100.7 &  471  &   2 & 
60,L     & 1.3 &  0.81 & 113,L &
107,96,71,38,60,39,113 \\
  455.0 & 0.5427 &   4.7 &   40 & 105,L    &   3.2 &   27  &  19 & 
12,X     & 2.7 &  0.73 & 12,X &
3,46,105,12,50,8,116,1,45 \\
  456.0 & 0.2330 &  12.0 &   60 & 41,L    &   8.1 &   40  &   1 & 41,C 
& 1.1* &  0.62 & 41,L & 107,41 \\
  458.0 & 0.290  & 232.2 & 1347 & 116,L    & 190.3 & 1104  &  14 & 
107,C    & 1.2 &  0.78 & 116,L & 107,116 \\
  459.0 & 0.2199 &  12.6 &   60 & 41,L    &   8.3 &   40  &   5 & 
116,U     & 5.0* &  0.65 & 41,L &
112,80,116,41,95 \\
  460.0 & 0.268  &   8.6 &   47 & 116,L    &   5.1 &   28  &   2 & 
116,U     & 5.0* &  0.60 & 116,L &
107,89,56,116,34,116 \\ \tableline
\end{tabular}
\normalsize
\tablecomments{(1) LAS and size values listed as lower limits were 
taken from maps that may be undersampled.
The LAS and size value listed as an upper limit are for an unresolved source.
\newline (2) Values marked with a colon were calculated using an 
optical position rather than a radio core
position.
\newline (3) For $Q$ values marked with an asterisk, the map 
reference and frequency band are those given in
column 9 rather than column 12.
\newline (4) Values of $FR$ listed as upper limits were taken from 
maps that may be undersampled.
\newline (5) Values of $FR$ that are marked with a dagger were 
computed with $l$ measurements that are
substantially smaller ($\lesssim 90\%$) than the distance to the 
farthest detectable emission.
\newline (6) The $FR$ value marked with a question mark may be
affected by poor map resolution.}
\end{table}
\begin{table}
\clearpage
\small
\tablerefs{1: Akujor \& Garrington 1995; 2: Akujor, Spencer \& 
Wilkinson 1990; 3: Akujor et~al.\ 1991; 4:
Alexander 1987; 5: P. Alexander 1994, private communication; 6: 
Alexander \& Leahy 1987; 7: Antonucci 1985; 8:
P.D. Barthel 1994, private communication; 9: S.A. Baum 1994, private 
communication; 10: Baum et~al.\ 1988; 11:
P.N. Best 1994, private communication; 12: Bogers et~al.\ 1994; 13: 
Bondi et~al.\ 1996; 14: Branson et~al.\
1972; 15: Bridle et~al.\ 1989; 16: Bridle et~al.\ 1994; 17: Bridle \& 
Perley 1984; 18: Bridle, Perley \&
Henriksen 1986; 19: Browne et~al.\ 1982a; 20: Browne et~al.\ 1982b; 
21: Burns et~al.\ 1984; 22: Clarke et~al.\
1992; 23: Clarke \& Burns 1991; 24: Conway et~al.\ 1993; 25: Downes 
et~al.\ 1986; 26: Dreher \& Feigelson
1984; 27: Fanti et~al.\ 1985; 28: Fanti et~al.\ 1989; 29: Feretti 
et~al.\ 1984; 30: Fernini, Burns \& Perley
1997; 31: Fernini et~al.\ 1991; 32: Garrington, Conway \& Leahy 1991; 
33: Giovannini, Feretti \& Gregorini
1987; 34: Giovannini et~al.\ 1988; 35: Giovannini et~al.\ 1994; 36: 
Gower \& Hutchings 1984; 37: Gregorini
et~al.\ 1988; 38: M.J. Hardcastle 1994, private communication; 39: 
Hardcastle et~al.\ 1997; 40: Hargrave
\& McEllin 1975; 41: Harvanek \& Hardcastle 1998; 42: Heckman, van Breugel
\& Miley 1984; 43: Hintzen, Ulvestad \& Owen 1983; 44: D.H. Hough 
1994, private communication; 45: Hutchings
et~al.\ 1998; 46: Jenkins, Pooley
\& Riley 1977; 47: Johnson, Leahy \& Garrington 1995; 48: W. Junor 
1994, private communication; 49: W. Junor
1995, private communication; 50: V.K. Kapahi 1994, private 
communication; 51: D.M. Katz-Stone 1994, private
communication; 52: D.M. Katz-Stone 1995, private communication; 53: 
Katz-Stone \& Rudnick 1997; 54: Kollgaard,
Wardle \& Roberts 1989; 55: Kronberg \& Reich 1983; 56: Laing 1981; 
57: Laing 1989; 58: J.P. Leahy 1994,
private communication; 59: Leahy 1996; 60: Leahy, Bridle \& Strom 
1997; 61: Leahy, Muxlow \& Stephens 1989;
62: Leahy \& Perley 1991; 63: Leahy, Pooley \& Riley 1986; 64: Leahy 
\& Williams 1984; 65: Liu
\& Pooley 1990; 66: Liu \& Pooley 1991a; 67: Liu \& Pooley 1991b; 68: 
Liu, Pooley
\& Riley 1992; 69: Lonsdale \& Morison 1983; 70: E. Ludke 1994, 
private communication; 71: Macdonald,
Kenderdine \& Neville 1968; 72: Mackay 1969; 73: F. Mantovani 1994, 
private communication; 74: Mantovani
et~al.\ 1992; 75: Mantovani et~al.\ 1997; 76: Mantovani et~al.\ 1990; 
77: McCarthy, van Breugel \& Kapahi
1991; 78: McCarthy, van Breugel \& Spinrad 1989; 79: Miller, Rawlings 
\& Saunders 1993; 80: Morganti, Killeen
\& Tadhunter 1993; 81: Myers \& Spangler 1985; 82: Neff, Roberts \& 
Hutchings 1995; 83: Pearson, Perley \&
Readhead 1985; 84: J.A. Pedelty 1994, private communication; 85: 
Pedelty et~al.\ 1989; 86: Perley et~al.\
1980; 87: Perley, Fomalont \& Johnston 1980; 88: Perley \& Taylor 
1991; 89: Pooley \& Henbest 1974; 90: Price
et~al.\ 1993; 91: Rantakyr\"{o}, B{\aa\aa}th \& Matveenko 1995; 92: 
Rantakyr\"{o} et~al.\ 1992; 93: Reid
et~al.\ 1995; 94: Rendong et~al.\ 1991; 95: Rhee et~al.\ 1996; 96: 
Riley \& Pooley 1975; 97: Rocca-Volmerange
et~al.\ 1994; 98: Rudnick \& Adams 1979; 99: Sanghera et~al.\ 1995; 
100: Schilizzi \& de Bruyn 1983; 101:
Simon et~al.\ 1990; 102: Spangler, Myers \& Pogge 1984; 103: Spangler 
\& Pogge 1984; 104: Spangler \& Sakurai
1985; 105: Spencer et~al.\ 1989; 106: Spencer et~al.\ 1991; 107: J.T. 
Stocke 1994, private communication; 108:
Stocke, Burns \& Christiansen 1985; 109: Strom et~al.\ 1990; 110: Swarup, Sinha
\& Hilldrup 1984; 111: Taylor \& Perley 1992; 112: Ulvestad 1985; 
113: Unknown; 114: Unwin \& Wehrle 1992;
115: W.J.M. van Breugel 1994, private communication; 116: W.J.M. van 
Breugel 1995, private communication; 117:
van Breugel et~al.\ 1992; 118: van Breugel, Miley \& Heckman 1984; 
119: Vigotti et~al.\ 1989; 120: Waak
et~al.\ 1988; 121: Wilkinson et~al.\ 1991; 122: Zensus, Cohen \& Unwin 1995.}
\end{table}

\begin{table}
\caption{EMSS Cluster $B_{gg}$ Values}
\label{emssbggs}
\tiny
\begin{tabular}{cccc}
\tableline \tableline
  EMSS  & $z$ & B$_{gg}$       & Comments \\ Cluster&     & (Mpc$^{1.77}$) & \\
\tableline
  MS0011.7+0837     & 0.163 &  1028.  &  \\
  MS0015.9+1609$^*$ & 0.546 &  2350.  & Yee \\
  MS0026.4+0725     & 0.170 &   708.  &  \\
  MS0109.4+3910     & 0.208 &   527.  &  \\
  MS0147.8-3941     & 0.373 &   703.  &  \\
  MS0159.1+0330     & 0.165 &   567.  &  \\
  MS0302.5+1717$^*$ & 0.425 &   564.  & Yee \\
  MS0302.7+1658$^*$ & 0.426 &  1163.  &  \\
  MS0353.6-3642     & 0.320 &  1188.  &  \\
  MS0418.3-3844     & 0.350 &   672.  &  \\
  MS0433.9+0957     & 0.159 &  1092.  &  \\
  MS0440.5+0204     & 0.190 &   906.  & Yee \\
  MS0451.5+0250     & 0.202 &  1275.  & Yee \\
  MS0451.6-0305$^*$ & 0.545 &  2664.  & Yee \\
  MS0735.6+7421     & 0.216 &   715.  & CF \\
  MS0810.5+7433     & 0.282 &   697.  &  \\
  MS0811.6+6301$^*$ & 0.312 &   725.  &  \\
  MS0821.5+0337$^*$ & 0.347 &   651.  &  \\
  MS0839.8+2938     & 0.194 &  1203.  & Yee; CF \\
  MS0849.7-0521     & 0.192 &   618.  &  \\
  MS0906.5+1110     & 0.180 &  1238.  &  \\
  MS1004.2+1238     & 0.166 &   554.  &  \\
  MS1006.0+1202     & 0.221 &  1554.  & Yee \\
  MS1008.1-1224$^*$ & 0.301 &  1758.  & Yee \\
  MS1020.7+6820     & 0.203 &   623.  &  \\
  MS1054.4-0321$^*$ & 0.823 &  1529.  &  \\
  MS1125.3+4324     & 0.181 &   507.  &  \\
  MS1137.5+6625$^*$ & 0.782 &  1400.  &  \\
  MS1147.3+1103$^*$ & 0.303 &   828.  &  \\
  MS1201.5+2824     & 0.167 &   782.  &  \\
  MS1208.7+3928$^*$ & 0.340 &   784.  &  \\
  MS1219.9+7542     & 0.240 &   641.  &  \\
  MS1224.7+2007$^*$ & 0.327 &   724.  & Yee \\
  MS1231.3+1542     & 0.238 &  1180.  & Yee; CF \\
  MS1241.5+1710$^*$ & 0.555 &  1500.  &  \\
  MS1244.2+7114     & 0.225 &  1036.  &  \\
  MS1253.9+0456     & 0.230 &   948.  &  \\
  MS1305.4+2941     & 0.241 &   672.  &  \\
  MS1308.8+3244     & 0.245 &   740.  &  \\
  MS1335.2-2928     & 0.189 &   643.  &  \\
  MS1358.4+6245$^*$ & 0.327 &  1601.  & Yee; CF \\
  MS1401.9+0437     & 0.230 &   541.  &  \\
  MS1409.9-0255     & 0.221 &   585.  &  \\
  MS1421.0+2955     & 0.261 &   662.  &  \\
  MS1426.4+0158$^*$ & 0.320 &  1020.  &  \\
  MS1455.0+2232     & 0.259 &   508.  & Yee; CF \\
  MS1512.4+3647$^*$ & 0.372 &   540.  & Yee; CF \\
  MS1532.5+0130$^*$ & 0.320 &   714.  &  \\
  MS1546.8+1132     & 0.226 &   921.  &  \\
  MS1617.1+3237     & 0.274 &   618.  &  \\
  MS1618.9+2552     & 0.161 &   817.  &  \\
  MS1621.5+2640$^*$ & 0.426 &  1496.  & Yee \\
  MS1910.5+6736     & 0.246 &  1099.  &  \\
  MS2053.7-0449$^*$ & 0.583 &  1239.  &  \\
  MS2137.3-2353     & 0.313 &  1917.  &  \\
  MS2142.7+0330     & 0.239 &   672.  &  \\
  MS2255.7+2039     & 0.288 &   786.  &  \\
  MS2301.5+1506     & 0.247 &   968.  &  \\
  MS2318.7-2328     & 0.187 &  1334.  &  \\ \tableline
\end{tabular}
\normalsize
\end{table}

%
\end{document}